\documentclass[sigconf]{acmart}

\AtBeginDocument{%
  }

\copyrightyear{2026}
\acmYear{2026}
\setcopyright{cc}
\setcctype{by}
\acmConference[CCS '26]{Proceedings of the 2026 ACM SIGSAC Conference on Computer and Communications Security}{November 15--19, 2026}{The Hague, Netherlands}
\acmBooktitle{Proceedings of the 2026 ACM SIGSAC Conference on Computer and Communications Security (CCS '26), November 15--19, 2026, The Hague, Netherlands}
\acmDOI{10.1145/3830454.3832684}
\acmISBN{979-8-4007-2871-6/2026/11}

\usepackage{multirow}
\usepackage{algorithm}
\usepackage{algorithmic}
\usepackage{subfig}
\usepackage{textcomp}
\usepackage{xspace}
\usepackage{makecell}
\usepackage{tabularx}
\usepackage{hhline}
\usepackage{siunitx}
\usepackage{fp}
\usepackage{threeparttable}

\newcommand{\etal}{{\rm et al. }}
\newcommand{\customtinysize}{\fontsize{8}{10}\selectfont}

\newcolumntype{C}{>{\centering\arraybackslash\hsize=.5\hsize\linewidth=\hsize}X}

\newcommand{\stripminus}[1]{\expandafter\the\numexpr-#1}

\newcommand{\dataset}{\textsc{HiGraph}\xspace}
\newcommand{\datasetDrebin}{\textsc{Drebin}\xspace}
\newcommand{\model}{\textsc{Hydra}\xspace}
\newcommand{\adda}{\textsc{Adda}\xspace}

\begin{document}

\title{HYDRA: Proactive Android Malware Drift Adaptation via Hierarchical Graph Contrastive Learning}
\titlenote{This is the author's version of the paper accepted at the 2026 ACM
SIGSAC Conference on Computer and Communications Security (CCS '26). It
includes the full appendix that was shortened in the proceedings version due
to space constraints. The Version of Record will be available at
\url{https://doi.org/10.1145/3830454.3832684}.}

\author{Han Chen}
\orcid{0000-0002-8474-6867}
\affiliation{%
  \institution{University of Technology Sydney}
  \city{Sydney}
  \state{New South Wales}
  \country{Australia}
}
\email{han.chen-7@student.uts.edu.au}

\author{Hanchen Wang}
\orcid{0000-0003-3158-9586}
\affiliation{%
  \institution{University of Technology Sydney}
  \city{Sydney}
  \state{New South Wales}
  \country{Australia}
}
\email{hanchen.wang@uts.edu.au}

\author{Hongmei Chen}
\orcid{0000-0002-4054-3654}
\authornote{Corresponding author.}
\affiliation{%
  \institution{Yunnan University}
  \city{Kunming}
  \state{Yunnan}
  \country{China}
}
\email{hmchen@ynu.edu.cn}

\author{Lu Qin}
\orcid{0000-0001-6068-5062}
\affiliation{%
  \institution{University of Technology Sydney}
  \city{Sydney}
  \state{New South Wales}
  \country{Australia}
}
\email{Lu.Qin@uts.edu.au}

\author{Wenjie Zhang}
\orcid{0000-0001-6572-2600}
\affiliation{%
  \institution{University of New South Wales}
  \city{Sydney}
  \state{New South Wales}
  \country{Australia}
}
\email{wenjie.zhang@unsw.edu.au}

\author{Ying Zhang}
\orcid{0000-0002-2674-1638}
\affiliation{%
  \institution{University of Technology Sydney}
  \city{Sydney}
  \state{New South Wales}
  \country{Australia}
}
\email{ying.zhang@uts.edu.au}

\renewcommand{\shortauthors}{Chen et al.}

\begin{abstract}
  Concept drift, driven by the rapid evolution of Android malware, severely degrades the performance of machine learning detectors. Current adaptation strategies are often reactive, responding only after performance has dropped and imposing a significant manual annotation burden, or they are proactive but rely on unstable adversarial training and incomplete, single-level graph representations. To overcome these limitations, we propose \model (\textbf{Hy}brid \textbf{Dr}ift \textbf{A}daptation), a proactive adaptation framework that learns drift-invariant representations from hierarchically structured data. \model first models applications using a hybrid graph structure, combining fine-grained Control Flow Graphs (CFGs) and coarse-grained Function Call Graphs (FCGs) to capture comprehensive behavioral patterns. It then introduces a novel cross-domain contrastive learning objective that aligns historical (source) and new (target) data distributions. By generating pseudo-labels for unlabeled target samples, our method pulls representations of semantically similar applications together, regardless of their domain, within a single, stable optimization process. This approach unifies feature learning and domain alignment, eliminating the need for complex adversarial objectives. Extensive experiments on large-scale, time-ordered malware datasets demonstrate that \model achieves substantially lower False Negative and False Positive Rates than state-of-the-art baselines while requiring up to 87.5\% fewer labeled samples. Our work thus offers a robust and efficient solution to combat concept drift in security applications.
\end{abstract}



\maketitle

\section{Introduction}

The Android ecosystem continues to be a primary target for malicious actors, with millions of new malware samples emerging annually, posing a significant threat to user privacy and device security~\cite{zhangEnhancingStateoftheartClassifiers2020a, liuDeepLearningAndroid2023}. A fundamental challenge in developing robust defenses is \textit{concept drift}, the phenomenon where malware statistics shift over time, causing the performance of machine learning detectors to degrade rapidly. Recent empirical studies demonstrate the severity of this challenge: commercial ML-based classifiers experience detection rate drops from nearly 100\% to below 80\%, or even to 60\%, within just three months~\cite{zhangEnhancingStateoftheartClassifiers2020a}, while research systems show F1 score deterioration from 0.99 to 0.76 after only 6 months of deployment~\cite{chenContinuousLearningAndroid2023}. Unlike natural data drift, malware evolution is intentionally adversarial; attackers deliberately employ techniques like obfuscation, packing, and code virtualization to create new variants that evade signature-based and machine learning-based systems~\cite{liRevisitingConceptDrift2025, biggioEvasionAttacksMachine2013}.

\begin{figure}[t]
  \centering
  \includegraphics[width=\linewidth, clip, trim=5 5 5 5]{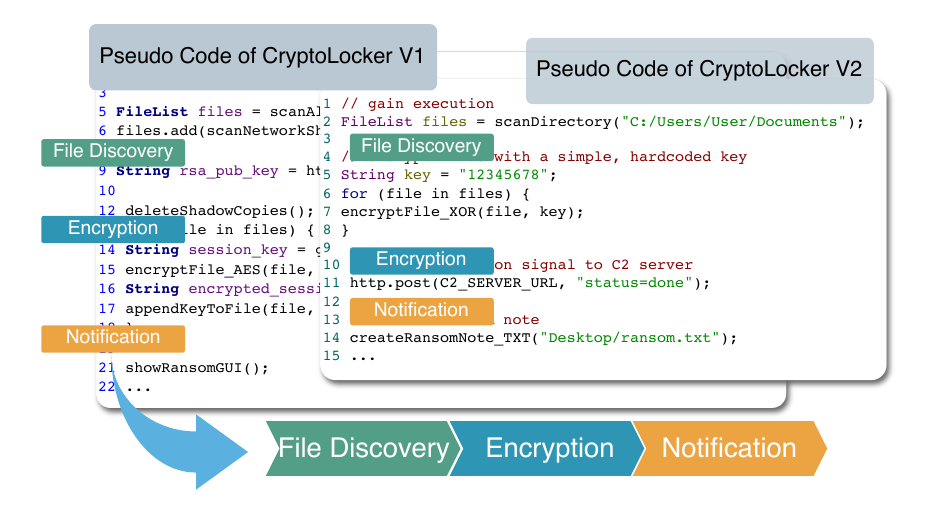}
  \caption{The behavioral evolution of CryptoLocker ransomware. The core malicious pattern (file discovery $\rightarrow$ encryption $\rightarrow$ notification) is preserved across versions despite significant implementation changes.}
  \label{fig:motivation}
  \vspace{-10pt}
\end{figure}

A critical insight underlying our work is that despite this continuous surface-level evolution, the \textbf{core malicious behavior} within a malware family often remains stable (see Figure~\ref{fig:motivation}). Malicious operations, such as stealing credentials, encrypting files for ransom, or exfiltrating data, depend on specific sequences of API calls and program structures that are fundamental to their goals. For example, a ransomware sample must interact with the file system and cryptographic APIs, a pattern that persists across variants. Graph-based representations of software are exceptionally well-suited to capture these behavioral invariants~\cite{bilotSurveyMalwareDetection2023,heMsDroidIdentifyingMalicious2023,zhangSemanticpreservingReinforcementLearning2022}. Control Flow Graphs (CFGs) can model the detailed, intra-procedural logic of individual functions, while Function Call Graphs (FCGs) can represent the high-level, inter-procedural interactions that constitute the overall malicious workflow~\cite{yangCADEDetectingExplaining2021, zhangEnhancingStateoftheartClassifiers2020a, zhangSemanticsAwareAndroidMalware2014,loGraphNeuralNetworkbased2022}. Crucially, such structural patterns are harder for adversaries to mutate than the flat static features (e.g., permissions, API strings) used by detectors like Drebin~\cite{arp2014drebin}, which can be altered with minimal effort.

Current strategies for handling malware drift, however, have significant limitations. The first category, \textbf{reactive adaptation}, responds to drift only after it is detected. CADE~\cite{yangCADEDetectingExplaining2021}, Dream~\cite{he2024combating}, and TRANSCENDENT~\cite{barberoTranscendingTranscendRevisiting2022} are notable examples of this paradigm. These methods rely on detecting drifting or out-of-distribution samples to trigger remediation, such as rejecting uncertain predictions or selecting informative samples for manual labeling to update the detector. While more efficient than random sampling, these methods are fundamentally reactive, as they act only after performance has already degraded. While recent works have proposed solutions for anticipating drift to trigger early retraining~\cite{tripathi2025towards}, the fundamental bottleneck remains the acquisition of reliable labels for this retraining process.
Furthermore, they impose a heavy burden on human experts, often requiring the manual inspection of more than $100$ samples per adaptation cycle, which is unsustainable at scale.

The second category, \textbf{proactive adaptation}, attempts to learn drift-invariant features upfront to build more resilient models. Recent pioneering work has applied domain adaptation techniques to this problem, using adversarial learning on CFGs to align feature distributions between older (source) and newer (target) malware populations~\cite{liRevisitingConceptDrift2025}. While conceptually powerful, this approach faces two critical hurdles. First, adversarial training is notoriously complex and unstable, requiring careful balancing of a generator and a discriminator in a min-max game. Second, and more critically, these methods rely on a single-level graph representation (CFGs), which captures local code patterns but fails to model the broader, inter-procedural call sequences and architectural patterns that are often the most stable indicators of a malware family's identity~\cite{lingMalGraphHierarchicalGraph2022}.

\begin{figure}[t]
  \centering
  \includegraphics[width=\linewidth]{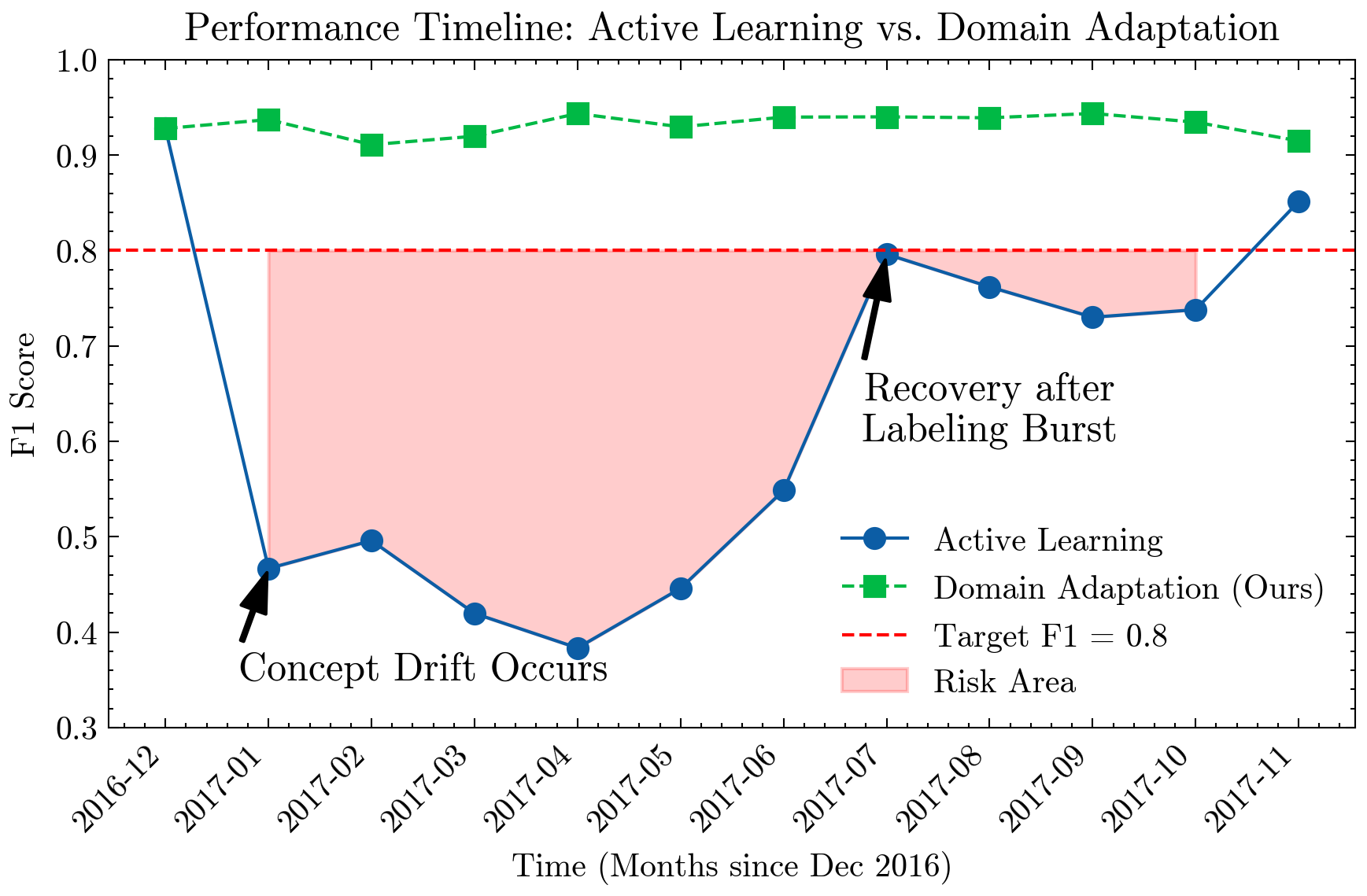}
  \caption{Reactive vs. Proactive Adaptation. Reactive methods (bottom) suffer from performance decay, requiring costly bursts of manual labeling to recover. Our proactive approach, \model~(top), continuously adapts by aligning distributions using hierarchical graphs. This maintains a high F1-score, reducing labeling effort by over 70\% and shrinking the ``risk area'' (time below performance threshold).}
  \label{fig:proactive_vs_reactive}
  \vspace{-10pt}
\end{figure}

This paper introduces \textbf{Hierarchical Graph Contrastive Learning} (\model), a novel domain adaptation framework that proactively learns drift-invariant representations for malware detection by addressing the limitations of prior works.
\model makes two primary technical innovations. First, we propose a \textbf{hierarchical graph representation}~\cite{lingMalGraphHierarchicalGraph2022, wangHierarchicalGraphBasedNeural2021,wang2020gognn, chen2023denoising} that integrates fine-grained, intra-procedural CFGs with a coarse-grained, inter-procedural FCG for each application.
This multi-level structure captures a more complete ``behavioral essence'' of malware, from the logic within a single function to the orchestration across the entire application.
Second, we design a \textbf{cross domain contrastive learning}~\cite{liuFewMHGCLFewShotMalware2024} framework. Instead of relying on complex adversarial optimization, our approach uses a more direct and stable contrastive objective. It effectively pulls representations of behaviorally similar malware (e.g., two variants of the same family from different time periods) closer together in the embedding space, while simultaneously pushing them apart from benign or dissimilar malware samples. This elegantly achieves both domain alignment and class separation in a unified, end-to-end process.

Our approach yields a more effective, efficient, and robust solution for adapting to malware concept drift. We summarize our key contributions as follows:
\begin{itemize}
  \item \textbf{Novel Proactive Adaptation Framework:} We propose \model, the first framework to combine hierarchical graph structures with cross-domain contrastive learning. Unlike reactive methods that wait for performance decay, \model continuously aligns source and target distributions to \emph{prevent} degradation rather than respond to it.

  \item \textbf{Hierarchical Behavioral Modeling:} We introduce a principled hierarchical graph representation that models both intra-procedural (CFG) and inter-procedural (FCG) behavior. This allows for the capture of more complete and stable malicious patterns that are missed by single-level graph approaches.

  \item \textbf{Operational Efficiency under Strict Labeling Budgets:} We demonstrate that \model substantially reduces both False Negative and False Positive Rates against state-of-the-art baselines while requiring up to 87.5\% fewer labeled samples. This directly addresses the label-acquisition bottleneck that dominates the operational cost of reactive drift adaptation.

  \item \textbf{Comprehensive Empirical Validation:} We conduct a thorough empirical study, including ablation analyses and parameter sensitivity tests, to validate the effectiveness of each component of our framework and demonstrate its robustness.
\end{itemize}

\noindent \textbf{Roadmap.} This paper is organized as follows: Section~\ref{sec:related} reviews related work. Section~\ref{sec:methodology} details our \model framework. Section~\ref{sec:real_world_drift} presents our experimental results. Finally, we present a discussion and conclusions.

\section{Related Work}
\label{sec:related}

The performance of malware classifiers degrades over time due to concept drift, the distributional shift in data caused by evolving threats~\cite{fanHeterogeneousTemporalGraph2021,gaoComprehensiveStudyLearningbased2024,arpDosDontsMachine2022}. To address this, we propose a domain adaptation framework that learns drift-invariant representations from hierarchical program graphs using contrastive learning. This section situates our work relative to prior art in concept drift, learning paradigms for adaptation, graph-based security analysis, and contrastive learning in cybersecurity.

\subsection{Concept Drift in Malware Detection}

The rapid evolution of malware causes distributional shifts that degrade detection performance. Li~\etal~\cite{liRevisitingConceptDrift2025} analyze this problem, showing that conventional retraining from scratch or via fine tuning is insufficient for adapting to drift with limited labels. Their work pioneers adversarial domain adaptation for malware, but its reliance on single-level Control Flow Graphs (CFGs) limits its ability to capture hierarchical program structures. Chen~\etal~\cite{chenContinuousLearningAndroid2023} employ hierarchical contrastive learning to build a similarity based uncertainty metric for continuous learning on Android malware. While effective against class imbalance, their method is reactive, requiring periodic retraining, and overlooks the rich information within program graph structures. Similarly, CADE~\cite{yangCADEDetectingExplaining2021} uses a contrastive autoencoder to embed behavioral concepts into a latent space for drift detection and explanation. This reduces labeling effort but, like others, it does not exploit the hierarchical structure of program graphs for more robust representations. Other work has focused on API semantics to improve robustness against malware evolution~\cite{zhangEnhancingStateoftheartClassifiers2020a}. While effective, this line of research often overlooks the deeper structural relationships captured by program graphs.

\subsection{Proactive Domain Adaptation versus Reactive Detection}

Prior work typically addresses concept drift through drift detection mechanisms that support reactive adaptation strategies. Methods like CADE~\cite{yangCADEDetectingExplaining2021} and Transcend~\cite{barberoTranscendingTranscendRevisiting2022,jordaneyTranscendDetectingConcept} focus on identifying drifting examples, serving either as query strategies within an active learning loop or as decision criteria for classification-with-rejection. While these approaches effectively flag out-of-distribution samples, they remain inherently reactive: the system must first detect drift before triggering remediation steps, such as manual labeling for retraining or rejecting uncertain predictions. This reactive posture limits the ability to learn representations that are inherently robust to malware evolution. In contrast, domain adaptation~\cite{long2015learning} offers a proactive solution by learning features that are invariant across time periods. While transfer learning has shown promise, it has not been combined with hierarchical representations and contrastive objectives for this task. Li \etal ~\cite{liRevisitingConceptDrift2025} made an early attempt at adversarial domain adaptation~\cite{ganin2016domain} using single level CFGs. Our work advances this proactive paradigm by developing a domain adaptation framework specifically for hierarchical graph representations, capturing both fine grained intra-procedural and coarse grained inter-procedural patterns.

\subsection{Graph-Based Security Analysis}

Program graphs are powerful representations for malware analysis as they capture structural patterns resistant to obfuscation~\cite{freitasLargeScaleDatabaseGraph2021,zhangEnhancingStateoftheartClassifiers2020a,liuDeepLearningAndroid2023}. Control Flow Graphs (CFGs) model intra procedural execution flow, while Function Call Graphs (FCGs) capture inter procedural interactions, offering complementary views of program behavior~\cite{gaoGDroidAndroidMalware2021}. Although Graph Neural Networks (GNNs)~\cite{kipf2016semi,xu2018powerful,velivckovic2017graph} are widely used for malware detection on program graphs, they suffer from two key limitations. First, most approaches operate on single level graphs (e.g., only CFGs or FCGs) or use high level semantic graphs~\cite{zhangEnhancingStateoftheartClassifiers2020a}, failing to capture the rich, multi level structure of software. Second, they often treat CFGs and FCGs as independent sources of information~\cite{liRevisitingConceptDrift2025,liBlackboxAdversarialExample2023}, missing the opportunity to learn unified representations that bridge local and global program behavior. Our work directly addresses this gap by introducing a hierarchical learning framework that jointly models both graph types.

\subsection{Contrastive Learning for Cybersecurity}

Contrastive learning is effective for security tasks because it learns robust similarity measures. For instance, it has been used to handle class imbalance in malware detection~\cite{chenContinuousLearningAndroid2023}, for dynamic analysis~\cite{yang2022android}, and to learn conceptual embeddings for drift detection~\cite{yangCADEDetectingExplaining2021}. However, these methods typically apply contrastive objectives to flat feature vectors~\cite{liuFewMHGCLFewShotMalware2024} or single level graph structures~\cite{liBlackboxAdversarialExample2023}. They do not exploit the natural hierarchy of program representations for domain adaptation~\cite{liRevisitingConceptDrift2025}. Our work is the first to design a contrastive learning framework that operates across a hierarchy of program graphs to learn domain invariant features.

\subsection{Positioning Our Approach}

Our work makes three primary contributions to the study of concept drift in malware detection. First, unlike reactive active learning methods, our proactive domain adaptation framework learns representations that are inherently robust to drift. Second, we introduce a hierarchical graph learning framework that captures both fine grained (CFG) and coarse grained (FCG) program structure, creating a more comprehensive view than single level graph approaches. Third, we are the first to apply contrastive learning directly to these hierarchical structural representations, learning a similarity metric that is robust to both code obfuscation and domain shift.
\section{Methodology}
\label{sec:methodology}

This section details our Hierarchical Graph Contrastive Learning (\model) framework, a graph-based approach for Android malware detection designed to address concept drift. We first introduce key concepts and problem formulation, then present our hierarchical graph construction and contrastive learning framework.

\subsection{Preliminaries}
\label{sec:preliminaries}

\textbf{Hierarchical Graph Representation.}
\label{sec:hg_representation}
To model an application's behavior, we construct a \textbf{hierarchical graph} that captures both intra-procedural logic and inter-procedural interactions. This structure consists of two levels: fine-grained Control Flow Graphs (CFGs) for each function, and a single, application-level Function Call Graph (FCG).

\textbf{Control Flow Graph (CFG).} For each function $f_j$ in the application, we extract a directed graph $G_{cfg}^{(j)} = (V_{cfg}^{(j)}, E_{cfg}^{(j)})$. The nodes $V_{cfg}^{(j)}$ are the basic blocks (straight-line code sequences), and the edges $E_{cfg}^{(j)}$ represent control flow between these blocks. Each node $v \in V_{cfg}^{(j)}$ is associated with a feature vector $\mathbf{x}_v \in \mathbb{R}^{d_{cfg}}$ that encodes its semantic properties (e.g., opcodes, API calls).

\textbf{Function Call Graph (FCG).} For the entire application, we build a single FCG, denoted $G_{fcg} = (V_{fcg}, E_{fcg})$. The nodes $V_{fcg}$ represent all functions $\{f_j\}_{j=1}^J$ in the application, and an edge $(f_i, f_j) \in E_{fcg}$ exists if function $f_i$ calls function $f_j$. Each node $f_j \in V_{fcg}$ is associated with a feature vector encoding its signature attributes.

\begin{figure}[t]
  \centering
  \includegraphics[width=0.9\linewidth]{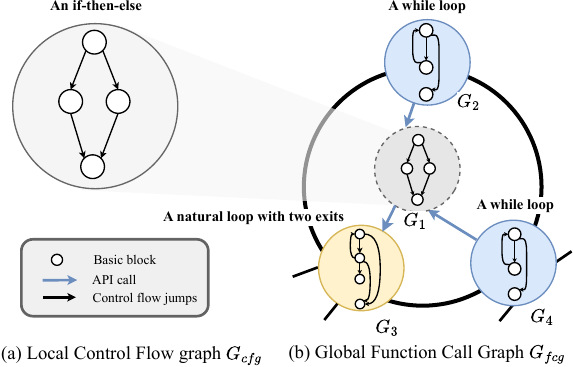}
  \caption{An illustration of our hierarchical graph representation. An application is modeled as a high-level Function Call Graph (FCG), where each node (a function) is internally represented by its own detailed Control Flow Graph (CFG).}
  \label{fig:hg_overview}
\end{figure}

The complete hierarchical representation for an application is the collection $G = (G_{fcg}, \{G_{cfg}^{(j)}\}_{j=1}^{J})$, which provides a comprehensive basis for our analysis. Figure~\ref{fig:hg_overview} illustrates this hierarchical structure.

\noindent\textbf{Operational Rationale.} This hierarchical representation is designed for robustness against concept drift. We hypothesize that high-level, inter-procedural workflows (captured by FCGs) remain more stable over time than low-level, intra-procedural implementation details (captured by CFGs). Our model leverages this asymmetry to learn drift-invariant features, as detailed in Section~\ref{sec:graph_learning}.

\noindent\textbf{Problem Formulation.}
\label{sec:problem} Let $\mathcal{D}_s = \{(G_i^s, y_i^s)\}_{i=1}^{n_s}$ be the labeled source domain, where $G_i^s$ is the hierarchical graph representation of an application and $y_i^s \in \{0, 1\}$ is its label (benign/malicious). Similarly, let $\mathcal{D}_t = \{(G_i^t, y_i^t)\}_{i=1}^{n_t}$ be the target domain containing newer applications. We operate under a realistic scenario where only a small, budgeted subset of target data is labeled.

Concept drift occurs when the joint data distribution shifts between domains, i.e., $P_s(G, Y) \neq P_t(G, Y)$. This shift renders models trained on $\mathcal{D}_s$ ineffective on $\mathcal{D}_t$. Our objective is to learn a function $\psi(G)$ that maps a hierarchical graph $G$ to a drift-invariant representation, enabling a classifier to maintain high performance on the target domain $\mathcal{D}_t$ with minimal labeling cost.

\subsection{Hierarchical Graph Representation Learning}
\label{sec:graph_learning}

In this section, we describe how our graph representation learning framework learns meaningful representations from this multi-level structure using a two-stage GNN encoder on the hierarchical graph structure in the preliminaries.

\noindent\textbf{Motivation.}
Our central hypothesis is that while the low-level implementation details of malware may change frequently due to adversarial evolution (e.g., obfuscation), the high-level malicious workflow often remains stable. For example, a ransomware family will consistently exhibit a pattern of file discovery, encryption, and notification, even if the specific functions used are altered. To capture this duality, our framework learns representations from the hierarchical graph. The CFG-level encoder focuses on the fine-grained, intra-procedural semantics that are prone to drift, while the FCG-level encoder captures the coarse-grained, inter-procedural architecture, which is more likely to be drift-invariant. By integrating these two views, the model can learn to identify the stable, high-level ``behavioral essence'' of malware, making it more robust to concept drift.

\noindent\textbf{Intra-Procedural (CFG-Level) Encoding.}
For each function $f_j$, we construct its CFG, $G_{cfg}^{(j)} = (V^{(j)}, E^{(j)})$, where nodes represent basic blocks, each with an initial feature vector $\mathbf{x}_v \in \mathbb{R}^{d_{cfg}}$. To derive a function-level embedding, we aggregate node features via a single message-passing step.
The matrix of aggregated node features $\mathbf{H}$ is computed as:
\begin{align}
  \mathbf{H} = \hat{\mathbf{D}}^{-1}\hat{\mathbf{A}}\mathbf{X}
\end{align}
where $\mathbf{X}$ is the matrix of initial node features, $\hat{\mathbf{A}} = \mathbf{A} + \mathbf{I}$ is the adjacency matrix of $G_{cfg}^{(j)}$ with added self-loops, and $\hat{\mathbf{D}}$ is the corresponding diagonal degree matrix. This operation averages the features of a node with those of its direct neighbors. Second, a mean pooling operation is applied across all node embeddings in $\mathbf{H}$ to produce the final function-level embedding $\mathbf{f}_j$:
\begin{align}
  \mathbf{f}_j = \frac{1}{|V^{(j)}|} \sum_{v \in V^{(j)}} \mathbf{h}_v
\end{align}
This embedding summarizes the granular execution semantics of the function.

\noindent\textbf{Inter-Procedural (FCG-Level) Encoding.}
Next, we construct an application-wide FCG, where nodes represent functions and edges denote inter-procedural calls. Node initialization depends on the function type. For an internal function $u$, its node is initialized with its CFG-level embedding $\mathbf{f}_u$. For an external API call that lacks a local CFG, we initialize its node using a semantic embedding of its function name.
An FCG-level encoder based on the Graph Isomorphism Network (GIN)~\cite{xu2018powerful} then processes the entire FCG to produce the final application representation $\mathbf{g}$. The update rule for a node $v$ at layer $k$ in GIN is given by:
\begin{align}
  \mathbf{h}_v^{(k)} = \text{MLP}^{(k)}\left((1 + \epsilon^{(k)})\mathbf{h}_v^{(k-1)} + \sum_{u \in \mathcal{N}(v)} \mathbf{h}_u^{(k-1)}\right)
\end{align}
where $\mathbf{h}_v^{(k)}$ is the feature vector of node $v$ at layer $k$, $\mathcal{N}(v)$ is the set of its neighbors, and $\epsilon^{(k)}$ is a learnable parameter. After $K$ layers of GIN updates, we obtain the final node representations $\{\mathbf{h}_v^{(K)}\}_{v \in V_{fcg}}$. The overall graph representation $\mathbf{g}$ is then produced by summing over all node representations:
\begin{align}
  \mathbf{g} = \sum_{v \in V_{fcg}} \mathbf{h}_v^{(K)}
\end{align}
This hierarchical process yields a representation $\mathbf{g} = \psi(G)$ that integrates both local code patterns and global application architecture.

\subsection{Hierarchical Graph Contrastive Learning (\model)}
\label{sec:hgcl_framework}

To address the distribution shift between the source and target domains, we introduce a cross-domain contrastive learning objective. Traditional domain adaptation methods often rely on complex and unstable adversarial training. In contrast, our approach provides a more direct and stable mechanism for domain alignment. The core idea is to learn a representation space where semantically similar samples are clustered together, regardless of whether they come from the source or target domain. By defining ``similarity'' based on class labels (both true and pseudo-labels), the contrastive loss simultaneously achieves two critical goals for drift adaptation: (1) it pulls representations of the same malware class from different time periods (domains) together, and (2) it pushes representations of different classes apart. This process naturally aligns the source and target distributions in a class-aware manner, learning features that are invariant to the temporal drift.

\begin{figure*}[t]
  \centering
  \includegraphics[width=0.9\linewidth]{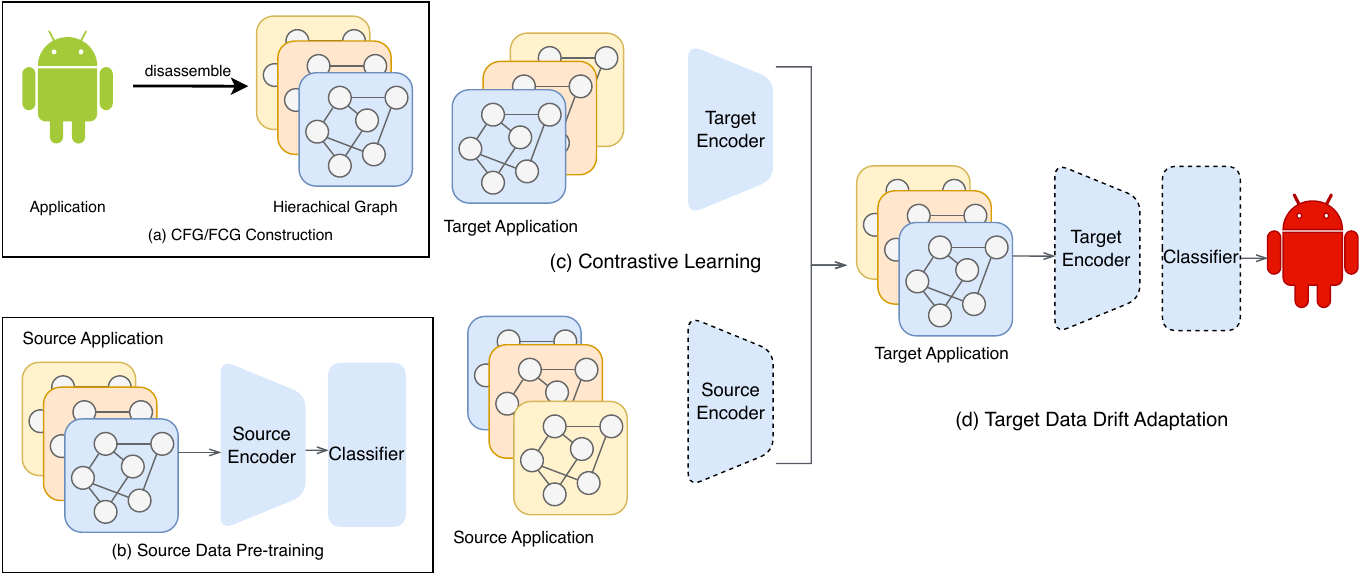}
  \caption{The \model framework architecture. The model learns drift-invariant representations by jointly optimizing a cross-domain contrastive loss and a supervised classification loss, effectively aligning source and target domains while preserving class separation.}
  \label{fig:framework}
\end{figure*}

Our core innovation is a semi-supervised contrastive learning framework that adapts the model to the target domain. As illustrated in Figure~\ref{fig:framework}, the process strategically leverages labeled source data, a small labeled target set, and a large unlabeled target set.

The training process involves the following key steps:

\noindent\textbf{Source Pre-training.} We first pre-train the GNN encoder $G_\theta$ and a classifier $C_\phi$ on the labeled source domain $\mathcal{D}_s$ using a standard cross-entropy loss, $\mathcal{L}_{pretrain}$. This provides a strong initial model for the next step.

\noindent\textbf{Target Data Partitioning and Pseudo-Labeling.} We partition the target domain $\mathcal{D}_t$ into a small, labeled subset $\mathcal{D}_{tl}$ (according to a budget, $k$) and a larger, unlabeled subset $\mathcal{D}_{tu}$. We then use the pre-trained model to generate pseudo-labels $\hat{y}^{tu}$ for all samples in $\mathcal{D}_{tu}$. This results in three distinct data groups: source data with true labels $(x^s, y^s)$, labeled target data with true labels $(x^{tl}, y^{tl})$, and unlabeled target data with pseudo-labels $(x^{tu}, \hat{y}^{tu})$.

\noindent\textbf{Joint Optimization with a Unified Loss.} We jointly train the model on mini-batches sampled from all three groups. The total loss function combines two objectives:
\begin{align}
  \mathcal{L}_{total} = \mathcal{L}_{contrast} + \alpha \cdot \mathcal{L}_{cls}
  \label{eq:total_loss}
\end{align}
where $\alpha$ is a balancing hyperparameter.

The first term is a \textbf{cross-domain contrastive loss} ($\mathcal{L}_{contrast}$). For each anchor sample $\mathbf{h}_i$ in a combined batch drawn from source and target data, we define its positive set $\mathcal{P}_i$ to include all other samples in the batch that share the same class label (or pseudo-label). For instance, a source domain malware sample is pulled closer to not only other source malware from the same family but also target domain malware assigned the same pseudo-label. The negative set $\mathcal{N}_i$ consists of all samples with different labels. Formally, for an anchor with label $y_i$, the positive set is $\mathcal{P}_i = \{\mathbf{h}_p | y_p = y_i, p \neq i\}$ and the negative set is $\mathcal{N}_i = \{\mathbf{h}_n | y_n \neq y_i\}$. The InfoNCE-based loss is then:
\begin{align}
  \begin{split}
    \mathcal{L}_{contrast} = & -\frac{1}{|\mathcal{B}|} \sum_{i \in \mathcal{B}} \log \Biggl(                                                                                                                                                                \\
                             & \frac{\sum_{p \in \mathcal{P}_i} \exp(\mathbf{h}_i^T \mathbf{h}_p / \tau)}{\sum_{p \in \mathcal{P}_i} \exp(\mathbf{h}_i^T \mathbf{h}_p / \tau) + \sum_{n \in \mathcal{N}_i} \exp(\mathbf{h}_i^T \mathbf{h}_n / \tau)} \Biggr)
  \end{split}
  \label{eq:contrastive_loss}
\end{align}
where $\tau$ is a temperature parameter. This loss pulls same-class embeddings together, regardless of their domain, thereby achieving a class-aware alignment of the source and target representation spaces.

The second term is a \textbf{supervised classification loss} ($\mathcal{L}_{cls}$). Critically, this cross-entropy loss is computed \textit{only} on samples with true labels from both the source mini-batch ($\mathcal{B}_s$) and the labeled target mini-batch ($\mathcal{B}_{tl}$):
\begin{align}
  \begin{split}
    \mathcal{L}_{cls} = \frac{1}{2} \Biggl( & \frac{1}{|\mathcal{B}_s|}\sum_{i \in \mathcal{B}_s} \text{CE}(y_i^s, C(\mathbf{h}_i^s))                       \\
                                            & + \frac{1}{|\mathcal{B}_{tl}|}\sum_{j \in \mathcal{B}_{tl}} \text{CE}(y_j^{tl}, C(\mathbf{h}_j^{tl})) \Biggr)
  \end{split}
  \label{eq:cls_loss}
\end{align}
By excluding pseudo-labeled samples from this loss, we isolate the classifier from potential pseudo-label noise, ensuring its robustness. This unified objective allows for end-to-end optimization of both the encoder and the classifier.

Algorithm~\ref{alg:hgcl} summarizes the complete training procedure.

\begin{algorithm}[tb]\small
  \caption{\model Training Algorithm}
  \label{alg:hgcl}
  \begin{algorithmic}[1]
    \REQUIRE Source data $\mathcal{D}_s$, target data $\mathcal{D}_t$, GNN encoder $G_\theta$, classifier $C_\phi$;\\
    Labeling budget $k$, hyperparameters $\alpha, \tau$, learning rate $\eta$.
    \ENSURE Trained model parameters $\theta, \phi$.

    \STATE \textbf{// Phase 1: Pre-training and Data Partitioning}
    \STATE Pre-train $G_\theta, C_\phi$ on $\mathcal{D}_s$ with cross-entropy loss.
    \STATE Partition $\mathcal{D}_t$ into a labeled set $\mathcal{D}_{tl}$ of size $k$ and an unlabeled set $\mathcal{D}_{tu}$.
    \STATE For each sample $G_i \in \mathcal{D}_{tu}$, generate pseudo-label $\hat{y}_i \leftarrow C_\phi(G_\theta(G_i))$.
    \STATE
    \STATE \textbf{// Phase 2: Joint Adaptation via Contrastive Learning}
    \FOR{each training iteration}
    \STATE Sample mini-batches $\mathcal{B}_s \subset \mathcal{D}_s$, $\mathcal{B}_{tl} \subset \mathcal{D}_{tl}$, and $\mathcal{B}_{tu} \subset \mathcal{D}_{tu}$.
    \STATE Construct combined batch $\mathcal{B} \leftarrow \mathcal{B}_s \cup \mathcal{B}_{tl} \cup \mathcal{B}_{tu}$.
    \STATE Generate embeddings $\{\mathbf{h}_i \leftarrow G_\theta(G_i)\}_{i \in \mathcal{B}}$.
    \STATE Compute $\mathcal{L}_{contrast}$ on all samples in $\mathcal{B}$ using Eq.~\eqref{eq:contrastive_loss}.
    \STATE Compute $\mathcal{L}_{cls}$ on labeled batches $(\mathcal{B}_s \cup \mathcal{B}_{tl})$ using Eq.~\eqref{eq:cls_loss}.
    \STATE $\mathcal{L}_{total} \leftarrow \mathcal{L}_{contrast} + \alpha \cdot \mathcal{L}_{cls}$.
    \STATE Update parameters: $(\theta, \phi) \leftarrow (\theta, \phi) - \eta \nabla_{(\theta, \phi)} \mathcal{L}_{total}$.
    \ENDFOR
    \RETURN Trained parameters $\theta, \phi$.
  \end{algorithmic}
\end{algorithm}

\subsection{Computational Complexity}
The computational complexity is dominated by the GNN forward pass, which is linear in the number of nodes and edges in the graphs, and the contrastive loss computation, which is quadratic in the batch size $B$. The overall complexity is $O(L \cdot (|V|+|E|) + B^2)$, where $L$ is the number of GNN layers. This is comparable to standard GNN training frameworks.

\section{Real-World Drift Evaluation}
\label{sec:real_world_drift}
In this section, we compare our method against two kinds of drift adaptation schemes:~1) active learning techniques from most recent works and~2) domain adaptation based methods.

\subsection{Dataset}
\label{sec:dataset}

\begin{table*}[t!]
  \centering
  \small
  \sisetup{group-separator={,}}
  \caption{Overview of the \dataset dataset statistics from 2012 to 2022.}
  \begin{tabular}{l*{11}{r}}
    \toprule
    \textbf{Year}    & \textbf{2012} & \textbf{2013} & \textbf{2014} & \textbf{2015} & \textbf{2016} & \textbf{2017} & \textbf{2018} & \textbf{2019} & \textbf{2020} & \textbf{2021} & \textbf{2022} \\
    \midrule
    Malicious (M)    & \num{5920}    & \num{5422}    & \num{5160}    & \num{5572}    & \num{5034}    & \num{5708}    & \num{5477}    & \num{5760}    & \num{4675}    & \num{1866}    & \num{67}      \\
    Benign (B)       & \num{50391}   & \num{47273}   & \num{47164}   & \num{50219}   & \num{48570}   & \num{48802}   & \num{46198}   & \num{47865}   & \num{41949}   & \num{20131}   & \num{758}     \\
    M+B              & \num{56311}   & \num{52695}   & \num{52324}   & \num{55791}   & \num{53604}   & \num{54510}   & \num{51675}   & \num{53625}   & \num{46624}   & \num{21997}   & \num{825}     \\
    M/(M+B)          & 10.51\%       & 10.29\%       & 9.86\%        & 9.99\%        & 9.39\%        & 10.47\%       & 10.60\%       & 10.74\%       & 10.03\%       & 8.48\%        & 8.12\%        \\
    Malware Families & \num{108}     & \num{154}     & \num{177}     & \num{222}     & \num{244}     & \num{232}     & \num{179}     & \num{170}     & \num{105}     & \num{69}      & \num{6}       \\
    \bottomrule
  \end{tabular}
  \label{tab:dataset_stats}
\end{table*}

We evaluate our approach on \dataset~\cite{chen2025higraph}, a large-scale dataset of \num{499981} Android applications from AndroZoo~\cite{allix2016androzoo} spanning 2012--2022.
Following the methodology in~\cite{chenContinuousLearningAndroid2023}, we use AVClass2~\cite{sebastian2020avclass2} with VirusTotal reports~\cite{virustotalVirusTotal2025} for labeling, resulting in \num{50661} malicious samples across 683 families and \num{449320} benign ones. Samples that AVClass2 labeled as SINGLETON, meaning no family name could be identified, were grouped into an `unknown` category.
To mitigate temporal and spatial bias~\cite{pendleburyTESSERACTEliminatingExperimental2019, zhangEnhancingStateoftheartClassifiers2020a}, we maintain a constant malicious-to-benign ratio of approximately 1:9 for each month's data.
Table~\ref{tab:dataset_stats} details the yearly distribution.

\noindent\textbf{Why we cut off at 2022.}
The main longitudinal evaluation is restricted to 2012--2022 because reliable malware labels become scarce beyond that point. Following HCC~\cite{chenContinuousLearningAndroid2023}, we mark a sample as malicious only when at least $15$ VirusTotal (VT) engines flag it; under this rule \dataset yields only $67$ malicious apps in $2022$ and fewer than $10$ in any subsequent year. Section~\ref{sec:relaxed_threshold} reports a supplementary evaluation that relaxes the threshold to VT$\geq 4$ for 2022--2025 (Table~\ref{tab:dataset_stats_relaxed}), at the cost of noisier labels.

\begin{figure}[b]
  \centering
  \includegraphics[width=\linewidth]{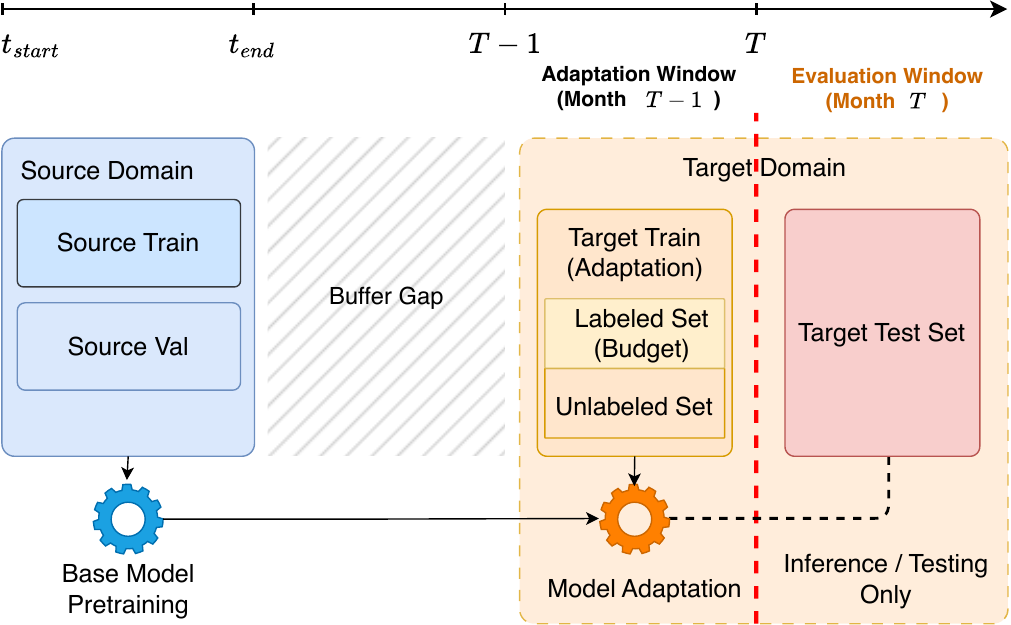}
  \caption{Overview of our time-aware rolling evaluation protocol. The model is pre-trained on a source domain, followed by a buffer gap to simulate realistic deployment delays where concept drift can occur. It is then evaluated in a rolling monthly window, adapting with data from month $T-1$ and testing on unseen data from the subsequent month $T$. This strict chronological separation mimics a realistic deployment scenario.}
  \label{fig:timeline}
\end{figure}

To create a realistic evaluation of concept drift, we adopt the time-consistent data split from~\cite{chenContinuousLearningAndroid2023, botacin2025towards}, forming two temporal adaptation tasks.
\textbf{Task A}: train on data from 2012 and test on 2013--2015.
\textbf{Task B}: train on data from 2016 and test on 2017--2022.
This chronological separation ensures that the model is always evaluated on future, unseen data, reflecting a real-world deployment scenario (Figure~\ref{fig:timeline}).

\subsection{Experimental Setup}
\label{sec:real_drift_setup}
To ensure a fair comparison, all methods are evaluated under an identical, time-consistent adaptation protocol. For both Task A (2012 training, 2013--2015 testing) and Task B (2016 training, 2017--2022 testing), models are first trained on the source year's data. Then, for each subsequent month $T$ in the test period, the model is adapted using a limited labeling budget drawn from the previous month's ($T-1$) data before being evaluated on the unseen samples of month $T$. This strict chronological separation ensures that the evaluation is performed on data that was not seen during the adaptation phase. This incremental, "warm-start" approach, where model weights from the previous month are carried over, is applied to all baselines as it is more effective for drift adaptation~\cite{chenContinuousLearningAndroid2023}.

\textbf{Graph Construction and Feature Extraction.} For all graph-based models (\model and ADDA), we use Androguard~\cite{desnos2018androguard} to decompile APKs and extract hierarchical graphs, comprising Function Call Graphs (FCGs) and Control Flow Graphs (CFGs). This process yields a large-scale dataset of over 200 million CFGs and nearly 600,000 FCGs. For CFG nodes, we extract an 11-dimensional feature vector based on instruction statistics within each basic block (e.g., counts of calls, transfers, arithmetic operations)~\cite{lingMalGraphHierarchicalGraph2022}. For FCG nodes corresponding to external API calls without a local CFG, we use a frozen, pre-trained API-name embedding to initialize node features (details in Appendix~\ref{sec:app-implementation}). For active learning baselines (TRANS, CADE, HCC), we extract the static feature set defined by Drebin~\cite{arp2014drebin} to ensure representation parity.

\textbf{Baselines.} Our focus is concept drift \emph{adaptation} rather than static detection: state-of-the-art static detectors degrade rapidly under distribution shifts and are unsuitable as longitudinal baselines. We compare our method against several state-of-the-art active learning and domain adaptation strategies, using the representations defined above.

\noindent\textit{Standard Classifiers.} We evaluate two standard classifiers: a linear Support Vector Machine (\textbf{SVM}) and a Multi-Layer Perceptron (\textbf{MLP}). Both models use the Drebin feature set~\cite{arp2014drebin} and follow the same incremental warm-start protocol as the other baselines to adapt to concept drift.

\noindent\textit{TRANS}~\cite{barberoTranscendingTranscendRevisiting2022} identifies drifted samples using credibility and confidence scores. We select samples with the lowest credibility scores for labeling.

\noindent\textit{CADE}~\cite{yangCADEDetectingExplaining2021} uses a contrastive autoencoder to learn embeddings and an Out-of-Distribution (OOD) score to identify drift. Samples with the highest OOD scores are selected. We enhance CADE by feeding its learned embeddings into a neural network classifier.

\noindent\textit{HCC}~\cite{chenContinuousLearningAndroid2023} is a prior state-of-the-art method that trains an encoder using a hierarchical contrastive loss and selects samples for labeling via a novel pseudo-loss uncertainty metric.

\noindent\textit{DREAM}~\cite{he2024combating} employs a model-sensitive contrastive autoencoder to detect drift and facilitates adaptation through human-in-the-loop revision of both family labels and behavioral explanations.

\noindent\textit{ADDA}~\cite{liRevisitingConceptDrift2025} is a well-established domain adaptation technique. For each adaptation step, the initial training set serves as the source domain, while the current month's unlabeled data serves as the target domain.

\noindent \textbf{\model Variants.} To isolate the benefits of continuous adaptation, we evaluate our model in three distinct operational modes: 1)~\textit{Static}, pre-trained once and used without adaptation; 2)~\textit{Adapt}, where the pre-trained model is reloaded and adapted each month; and 3)~\textit{Incremental}, our main proposal, which continuously adapts the model by carrying weights over from the previous month.

\noindent \textbf{Labeling Budget and Metrics.} To simulate realistic constraints, we evaluate performance with monthly labeling budgets of 50, 100, 200, and 400 samples. We report F1-Score, Accuracy, False Positive Rate (FPR), and False Negative Rate (FNR), emphasizing FNR due to its operational importance. All results are averaged across all test months for each task.

\noindent \textbf{Computational Cost.} Graph construction scales linearly at $\sim$100{,}000 APKs/hour, and each monthly adaptation completes within minutes using $\sim$24~GB peak memory on a single NVIDIA A40 (full hardware in Appendix~\ref{sec:app-implementation}).

\subsection{Results}
\label{sec:real_drift_results}

\begin{table*}[ht!]
  \centering
  \caption{Real drift adaptation on \dataset dataset (2012-2015 and 2016-2022) under different label budgets. Performance is measured by False Negative Rate (FNR), False Positive Rate (FPR), and F1-score. The best and second best results are in bold.}
  \resizebox{\textwidth}{!}{%
    \setlength{\tabcolsep}{1.5pt}
    \begin{tabular}{l l cccccc ccc cccccc ccc}
      \toprule
      \multirow{4}{*}{\textbf{Budget}} & \multirow{4}{*}{\textbf{Metric}} & \multicolumn{9}{c}{\textbf{\dataset (2013-2015)}} & \multicolumn{9}{c}{\textbf{\dataset (2017-2022)}}                                                                                                                                                                                                                                                                                                                                                                                                                                      \\
      \cmidrule(lr){3-11} \cmidrule(lr){12-20}
                                       &                                  & \multicolumn{6}{c}{\textbf{Active Learning}}      & \multicolumn{3}{c}{\textbf{Domain Adaptation}}    & \multicolumn{6}{c}{\textbf{Active Learning}} & \multicolumn{3}{c}{\textbf{Domain Adaptation}}                                                                                                                                                                                                                                                                                                                                      \\
      \cmidrule(lr){3-8}\cmidrule(lr){9-11}\cmidrule(lr){12-17}\cmidrule(lr){18-20}
                                       &                                  & \textbf{SVM}                                      & \textbf{MLP}                                      & \textbf{Trans.}                              & \textbf{CADE}                                  & \textbf{HCC}   & \textbf{DREAM} & \textbf{\adda} & {\footnotesize \textbf{\model\textsuperscript{†}}} & \textbf{\model} & \textbf{SVM} & \textbf{MLP} & \textbf{Trans.} & \textbf{CADE} & \textbf{HCC} & \textbf{DREAM} & \textbf{\adda}\textsuperscript{*} & {\footnotesize \textbf{\model\textsuperscript{†}}} & \textbf{\model} \\
      \midrule
      \multirow{3}{*}{50}
                                       & FNR                              & 61.12                                             & 64.95                                             & 48.49                                        & 46.45                                          & \textbf{40.46} & 54.46          & 60.09          & 47.95                                              & \textbf{38.72}
                                       & \textbf{34.32}                   & 34.73                                             & 51.30                                             & 51.51                                        & 38.19                                          & 44.42          & 54.39          & 34.42          & \textbf{32.84}                                                                                                                                                                                                                                                                  \\
                                       & FPR                              & 3.27                                              & 1.15                                              & \textbf{0.55}                                & 1.06                                           & 0.96           & 1.38           & \textbf{0.60}  & 1.60                                               & 0.75
                                       & 8.86                             & 4.14                                              & 1.85                                              & \textbf{1.42}                                & 2.44                                           & 1.52           & 2.29           & 1.87           & \textbf{0.62}                                                                                                                                                                                                                                                                   \\
                                       & F1                               & 89.84                                             & 90.81                                             & 66.24                                        & 65.36                                          & 71.27          & 92.17          & 91.82          & \textbf{92.86}                                     & \textbf{94.79}
                                       & 89.68                            & 93.01                                             & 50.46                                             & 58.79                                        & 60.47                                          & 93.66          & 92.76          & \textbf{94.59} & \textbf{95.83}                                                                                                                                                                                                                                                                  \\
      \midrule
      \multirow{3}{*}{100}
                                       & FNR                              & 61.24                                             & 61.92                                             & 44.50                                        & 43.95                                          & \textbf{38.92} & 50.32          & 53.56          & 46.37                                              & \textbf{37.60}
                                       & 34.12                            & 34.54                                             & 60.04                                             & 53.85                                        & 37.66                                          & 39.37          & 46.84          & \textbf{31.57} & \textbf{33.22}                                                                                                                                                                                                                                                                  \\
                                       & FPR                              & 3.06                                              & 1.12                                              & 1.01                                         & 1.07                                           & 1.12           & 1.47           & \textbf{0.67}  & 1.52                                               & \textbf{0.67}
                                       & 8.10                             & 3.78                                              & 1.80                                              & 1.40                                         & 2.40                                           & 1.51           & \textbf{0.92}  & 1.93           & \textbf{0.60}                                                                                                                                                                                                                                                                   \\
                                       & F1                               & 89.97                                             & 91.33                                             & 68.14                                        & 67.46                                          & 72.05          & 92.72          & 92.75          & \textbf{93.20}                                     & \textbf{95.00}
                                       & 90.21                            & 93.24                                             & 51.89                                             & 56.85                                        & 61.16                                          & 94.25          & 93.67          & \textbf{94.96} & \textbf{95.75}                                                                                                                                                                                                                                                                  \\
      \midrule
      \multirow{3}{*}{200}
                                       & FNR                              & 61.65                                             & 56.90                                             & 41.47                                        & 41.40                                          & \textbf{36.15} & 48.32          & 54.78          & 44.79                                              & \textbf{37.06}
                                       & 34.07                            & \textbf{32.96}                                    & 54.91                                             & 39.65                                        & 36.65                                          & 38.56          & 47.10          & \textbf{30.97} & 33.49                                                                                                                                                                                                                                                                           \\
                                       & FPR                              & 2.69                                              & 1.17                                              & \textbf{0.27}                                & 0.91                                           & 0.92           & 1.57           & \textbf{0.54}  & 1.58                                               & 0.71
                                       & 7.39                             & 3.66                                              & 2.75                                              & 1.87                                         & 2.76                                           & 1.93           & \textbf{0.94}  & 1.90           & \textbf{0.58}                                                                                                                                                                                                                                                                   \\
                                       & F1                               & 90.14                                             & 92.05                                             & 72.92                                        & 67.29                                          & 74.77          & 92.94          & 92.67          & \textbf{93.34}                                     & \textbf{95.00}
                                       & 90.68                            & 93.58                                             & 54.55                                             & 67.49                                        & 62.33                                          & 94.07          & 93.67          & \textbf{95.10} & \textbf{95.74}                                                                                                                                                                                                                                                                  \\
      \midrule
      \multirow{3}{*}{400}
                                       & FNR                              & 63.35                                             & 53.94                                             & 36.88                                        & 43.02                                          & \textbf{33.86} & 40.54          & 50.94          & 43.17                                              & \textbf{33.31}
                                       & 33.52                            & 29.12                                             & 54.20                                             & 41.12                                        & 35.71                                          & \textbf{26.74} & 38.52          & \textbf{27.74} & 28.39                                                                                                                                                                                                                                                                           \\
                                       & FPR                              & 2.33                                              & 1.06                                              & 1.20                                         & 1.02                                           & 1.01           & 1.72           & \textbf{0.64}  & 1.31                                               & \textbf{0.72}
                                       & 6.13                             & 2.75                                              & 2.28                                              & 1.80                                         & 2.80                                           & 1.33           & \textbf{0.78}  & 1.61           & \textbf{0.56}                                                                                                                                                                                                                                                                   \\
                                       & F1                               & 90.13                                             & 92.57                                             & 73.42                                        & 69.40                                          & 76.16          & \textbf{93.81} & 93.15          & \textbf{93.81}                                     & \textbf{95.51}
                                       & 91.51                            & 94.54                                             & 56.34                                             & 66.62                                        & 63.00                                          & \textbf{95.96} & 94.89          & 95.62          & \textbf{96.38}                                                                                                                                                                                                                                                                  \\
      \bottomrule
    \end{tabular}%
  }
  \begin{tablenotes}
    \item[] \textsuperscript{†} \model variant using Drebin features (MLP) instead of graph features, with the same adaptation pipeline.
    \item[] \textsuperscript{*} Due to OOM, results for ADDA on the 2017-2022 dataset are only averaged up to Feb 2021.
  \end{tablenotes}
  \label{tab:real_drift_adaptation}
\end{table*}

\begin{table}[t!]
  \centering
  \small
  \caption{Comparison of F1-Scores for different adaptation methods under real drift conditions from 2013 to 2015. Note that the \textit{Static} variant does not use any labeling budget, so its performance is constant across budget levels.}
  \begin{tabular}{ccccc}
    \toprule
    \textbf{Budget}      & \textbf{Variant}     & \textbf{2013}  & \textbf{2014}  & \textbf{2015}  \\
    \midrule
    -                    & \textit{Static}      & 91.48          & 90.14          & 86.57          \\
    \midrule
    \multirow{2}{*}{50}  & \textit{Adapt}       & 93.84          & 91.69          & 89.03          \\
                         & \textit{Incremental} & \textbf{95.69} & \textbf{94.84} & \textbf{93.92} \\
    \midrule
    \multirow{2}{*}{100} & \textit{Adapt}       & 94.35          & 92.14          & 88.75          \\
                         & \textit{Incremental} & \textbf{95.97} & \textbf{95.04} & \textbf{94.06} \\
    \midrule
    \multirow{2}{*}{200} & \textit{Adapt}       & 94.51          & 92.63          & 89.91          \\
                         & \textit{Incremental} & \textbf{95.98} & \textbf{94.94} & \textbf{94.17} \\
    \midrule
    \multirow{2}{*}{400} & \textit{Adapt}       & 95.23          & 93.86          & 93.03          \\
                         & \textit{Incremental} & \textbf{96.19} & \textbf{95.35} & \textbf{95.06} \\
    \bottomrule
  \end{tabular}
  \label{tab:real_drift_f1_scores}
  \vspace{-10pt}
\end{table}

\textbf{Overall Performance.}
Across both temporal tasks and all labeling budgets, our proposed method, \model, demonstrates a significant and consistent performance advantage over all baselines. As detailed in Table~\ref{tab:real_drift_adaptation}, the \textit{Incremental} variant, which continuously adapts, consistently achieves the highest F1-scores and the lowest False Negative Rates (FNR). For instance, in Task A with a budget of 100 labels, \model~(Incremental) reduces the FNR to 37.60\%, a 3.39\% relative improvement over the next-best active learning baseline, HCC (FNR 38.92\%). This performance gap is maintained across nearly all budget levels and tasks.

Table~\ref{tab:real_drift_f1_scores} further isolates the benefits of our continuous adaptation strategy by comparing the F1-scores of our three model variants in Task A. The \textit{Incremental} approach consistently surpasses both the \textit{Static} baseline (no adaptation) and the \textit{Adapt} variant, which re-initializes the model from the pre-trained state each month. This underscores the critical role of preserving learned knowledge across temporal windows; by carrying weights forward, the incremental model avoids catastrophic forgetting and adapts more effectively to evolving data distributions. For example, with a budget of 100 in 2015, the \textit{Incremental} model achieves an F1-score of 94.06\%, significantly outperforming the \textit{Adapt} (88.75\%) and \textit{Static} (86.57\%) variants.

\noindent \textbf{Strategy vs. Representation.}
Comparing \model\textsuperscript{†} (using Drebin features) with other baselines reveals the efficacy of our adaptation strategy.
Even with simple static features, our incremental approach consistently outperforms state-of-the-art baselines; for instance, \model\textsuperscript{†} achieves a 94.59\% F1-score in the 2017--2022 task (Budget 50).
This confirms that our learning strategy is inherently robust.
However, incorporating graph features (\model) yields a further boost, drastically reducing FNR (e.g., from 47.95\% to 38.72\% in Task A), demonstrating that hierarchical graphs capture behavioral patterns that static features miss during drift.

\begin{figure*}[t!]
  \centering
  \includegraphics[width=\textwidth]{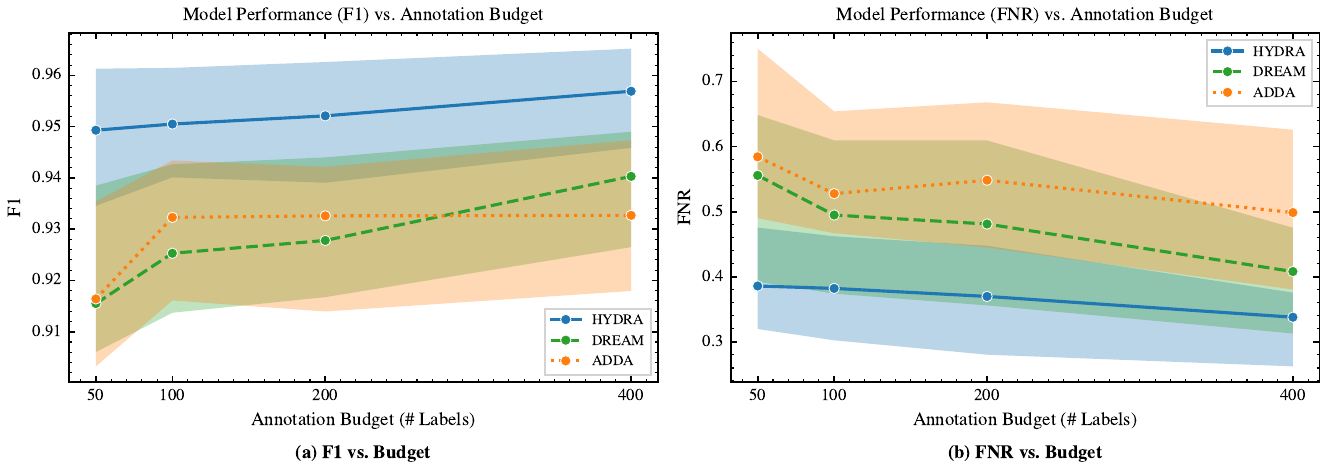}
  \caption{Performance vs. labeling budget. Our method achieves higher F1 and lower FNR with significantly less labeling budget compared to baselines. The shaded areas represent the 25-75\% quantile range of performance across all test months from 2013 to 2015, with the solid line indicating the median.}
  \label{fig:perf_vs_budget}
\end{figure*}

\noindent \textbf{Labeling Efficiency.}
Beyond superior performance, \model exhibits remarkable labeling efficiency.
Figure~\ref{fig:perf_vs_budget} illustrates that our method consistently operates on a more favorable performance-cost curve.
To achieve a better FNR than the state-of-the-art competitors (DREAM, ADDA), \model requires substantially fewer labels.
For example, our method with a budget of 50 achieves a median FNR of 0.39, while DREAM with 400 labels only reaches an FNR of 0.41. This represents an 87.5\% reduction in the number of required annotations, offering a substantial efficiency gain for security operations, even when accounting for the potentially varying costs of manual analysis.
This efficiency is critical in practice, enabling more frequent model updates and freeing up analyst resources.

\noindent\textbf{Labeling Cost Trade-offs.} \model achieves high performance with small labeling budgets (e.g., 50 samples/month), demonstrating significant efficiency gains over baselines. We note, however, that sample count is a simplified proxy for operational cost; in practice, manual analysis time varies by sample complexity. While our evaluation follows standard uniform-cost assumptions, future work could explore ``cost-aware'' adaptation that optimizes for analyst hours. Despite this nuance, the substantial reduction in required annotations (up to 87.5\%) indicates a clear practical benefit for reducing the burden on human experts.

\begin{figure*}[t!]
  \centering
  \subfloat[F1 - Incremental]
  {\includegraphics[width=0.48\textwidth]{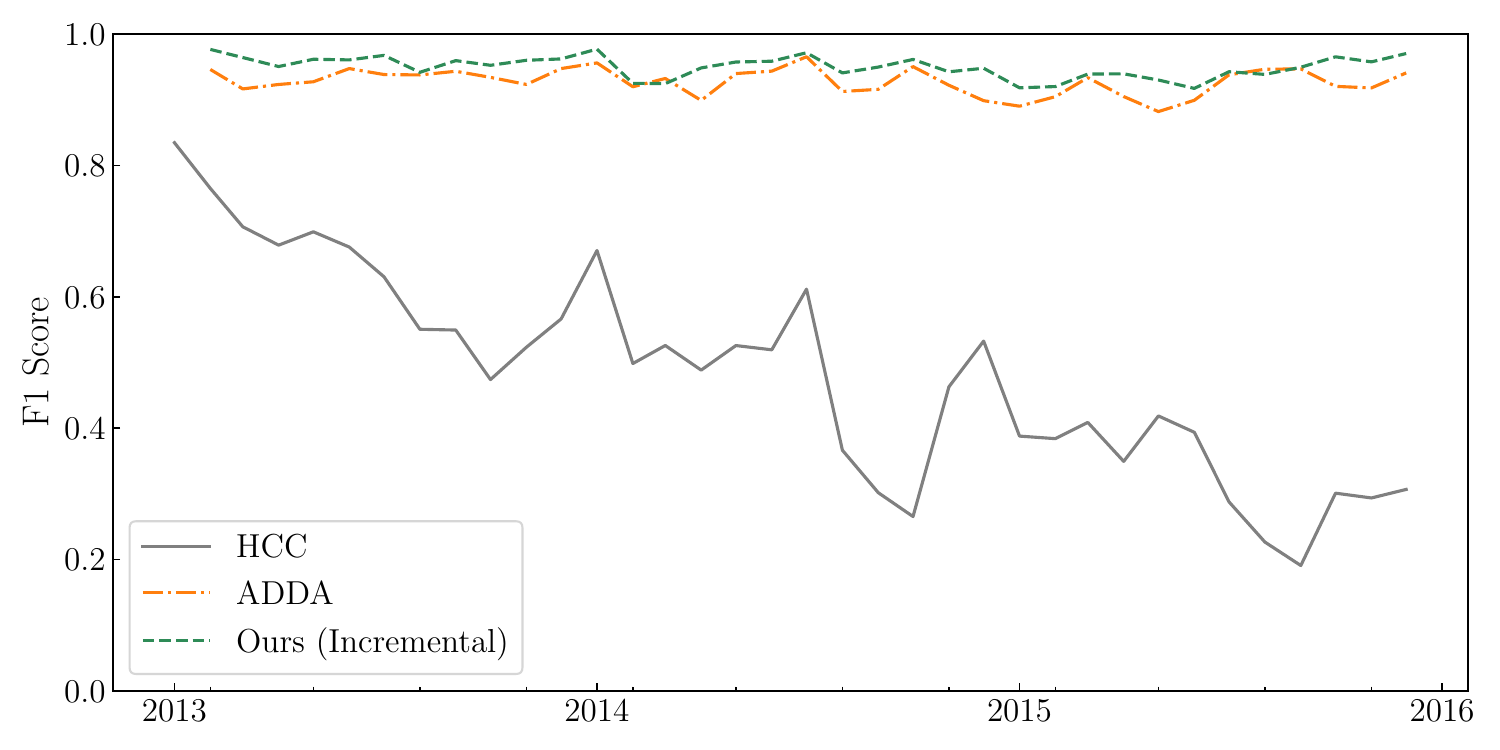}
    \label{fig:f1_incremental_hcc}}
  ~
  \subfloat[FNR - Incremental]
  {\includegraphics[width=0.48\textwidth]{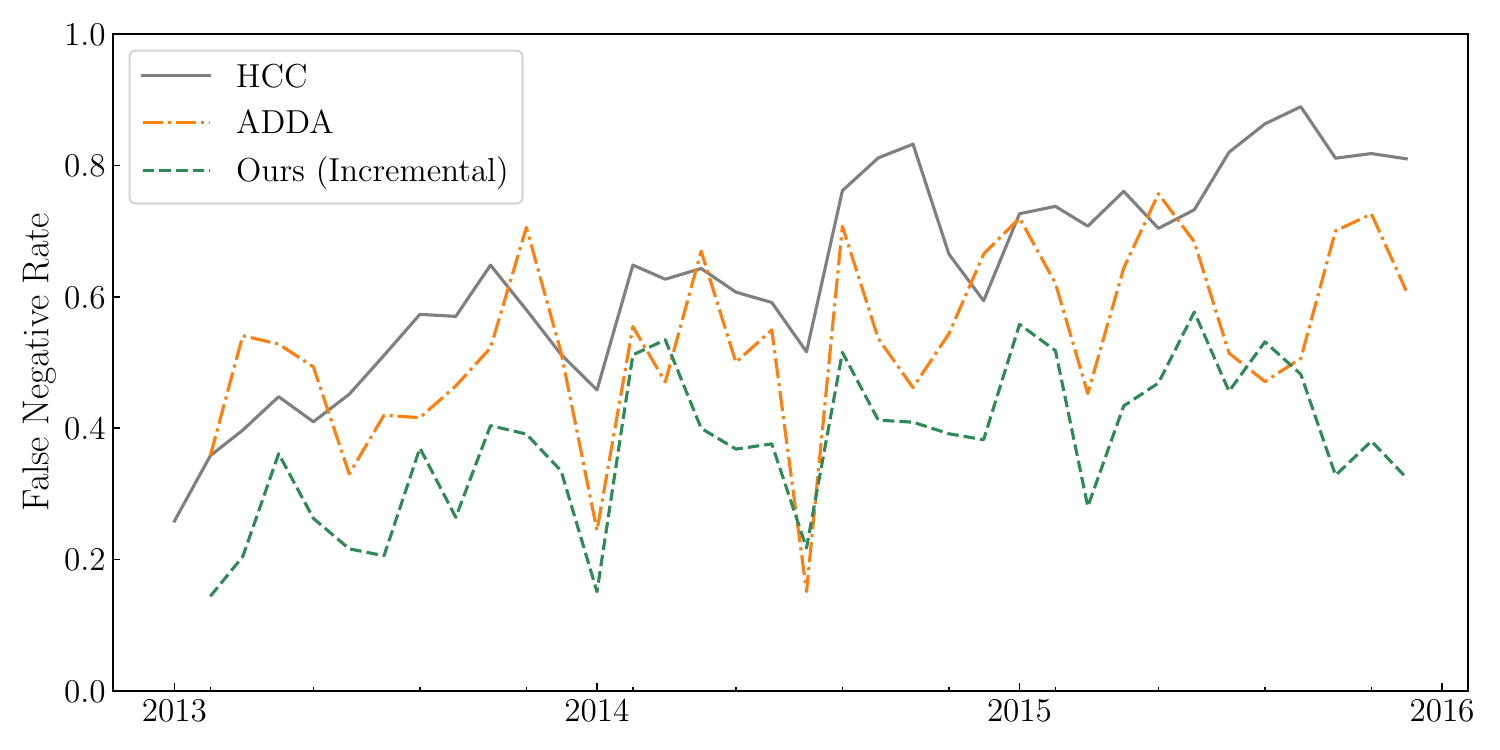}
    \label{fig:fnr_incremental_hcc}}
  \caption{Temporal performance. F1 and FNR of the incremental variant over time. The adaptation variant results are shown in Figure~\ref{fig:real_drift_hcc_adapt} in the Appendix.}
  \label{fig:real_drift_hcc_comparison}
\end{figure*}

\noindent\textbf{Robustness to Concept Drift.}
The sustained advantage of \model is demonstrated by its temporal performance, which shows resilience against concept drift.
Figure~\ref{fig:real_drift_hcc_comparison} plots the monthly F1 and FNR of our \textit{Incremental} variant against both HCC and ADDA; Figure~\ref{fig:real_drift_hcc_adapt} in Appendix~\ref{sec:app-results} provides the same view for the \textit{Adapt} variant. This granular, month-by-month view shows that our \textit{Incremental} approach consistently leads on both F1 and FNR over the entire test period, an insight not visible in the averaged results of Table~\ref{tab:real_drift_adaptation}. This visual evidence underscores our model's stability and consistent superiority over time.

\subsection{Relaxed-Label Evaluation on 2022--2025}
\label{sec:relaxed_threshold}

To evaluate \model's generalisability to recent malware, we extend our study to 2022--2025 AndroZoo applications. Because the strict VT$\geq 15$ threshold provides insufficient malicious samples during this period (Section~\ref{sec:dataset}), we relax it to VT$\geq 4$, yielding a 65{,}815-app dataset (Table~\ref{tab:dataset_stats_relaxed}). We maintain the same monthly rolling protocol, baselines, and hyperparameter settings as the main experiment (Section~\ref{sec:real_drift_setup}), using all of 2022 as the source domain.

\begin{table}[t!]
  \centering
  \small
  \sisetup{group-separator={,}}
  \caption{Dataset statistics for the relaxed VT$\geq$4 extension (2022--2025).}
  \label{tab:dataset_stats_relaxed}
  \begin{tabular}{lrrrrr}
    \toprule
    \textbf{Year}    & \textbf{2022} & \textbf{2023} & \textbf{2024} & \textbf{2025} & \textbf{Total} \\
    \midrule
    Malicious (M)    & \num{7607}    & \num{2086}    & \num{171}     & \num{21}            & \num{9885}     \\
    Benign (B)       & \num{38071}   & \num{16514}   & \num{1156}    & \num{189}           & \num{55930}    \\
    M+B              & \num{45678}   & \num{18600}   & \num{1327}    & \num{210}           & \num{65815}    \\
    M/(M+B)          & 16.65\%       & 11.22\%       & 12.89\%       & 10.00\%             & 15.02\%        \\
    Malware Families & 32            & 14            & 4             & 5                   & 39             \\
    \bottomrule
  \end{tabular}
\end{table}

\noindent\textbf{Performance under Relaxed Labels.}
As presented in Table~\ref{tab:relaxed_threshold_results}, \model achieves the highest F1-score at every labeling budget, peaking at 95.83\% with a budget of 400, and the lowest FNR at three of four budgets. Notably, while \adda attains the lowest FPR (0.20\%--0.68\%) across all budgets, its F1 trails \model by up to 3.4 percentage points, indicating that the FPR advantage comes at the cost of overall accuracy. \model therefore retains the best joint F1/FNR/FPR balance on the most recent (2023--2025) malware samples.

\noindent\textbf{Label-Noise Compression.}
The narrower 90--95\% F1 spread, far below the 50\%--96\% range under stricter VT$\geq 15$ labels in Table~\ref{tab:real_drift_adaptation}, reflects label noise that obscures methodological differences and justifies our VT$\geq 15$ choice for the main study.

\begin{table}[t!]
  \centering
  \caption{Adaptation results on the relaxed-threshold (VT$\geq$4) 2022--2025 dataset. Best and second-best per row are in \textbf{bold}.}
  \label{tab:relaxed_threshold_results}
  \resizebox{\linewidth}{!}{%
    \setlength{\tabcolsep}{3pt}
    \begin{tabular}{l l cccccc ccc}
      \toprule
      \multirow{2}{*}{\textbf{Budget}} & \multirow{2}{*}{\textbf{Metric}}
                                       & \multicolumn{6}{c}{\textbf{Active Learning}}
                                       & \multicolumn{3}{c}{\textbf{Domain Adaptation}}                                                                                                                                                                                                          \\
      \cmidrule(lr){3-8} \cmidrule(lr){9-11}
                                       &                                              & \textbf{SVM}    & \textbf{MLP}   & \textbf{Trans.}& \textbf{CADE}  & \textbf{HCC}  & \textbf{DREAM} & \textbf{\adda} & {\footnotesize \textbf{\model\textsuperscript{$\dagger$}}}  & \textbf{\model} \\
      \midrule
      \multirow{3}{*}{50}
                                       & FNR                                          & 9.62            & 9.54           & 9.55           & 9.17           & 9.89          & 10.86          & 10.58          & \textbf{8.29}                                               & \textbf{8.06}   \\
                                       & FPR                                          & 3.99            & 1.99           & 1.96           & 2.11           & 1.94          & 1.84           & \textbf{0.25}  & 2.28                                                        & \textbf{0.98}   \\
                                       & F1                                           & 90.84           & 92.92          & \textbf{92.99} & 92.78          & 92.89         & 92.80          & 92.01          & 92.68                                                       & \textbf{95.46}  \\
      \midrule
      \multirow{3}{*}{100}
                                       & FNR                                          & \textbf{8.76}   & 10.26          & 9.51           & 9.51           & 9.96          & 10.86          & 9.99          & \textbf{8.76}                                               & \textbf{8.64}   \\
                                       & FPR                                          & 3.98            & 1.89           & 2.04           & 2.04           & 1.90          & \textbf{1.85}  & \textbf{0.68}  & 2.08                                                        & 1.92            \\
                                       & F1                                           & 90.85           & 92.85          & 92.85          & 92.80          & 92.95         & 92.75          & \textbf{93.39} & 92.95                                                       & \textbf{93.87}  \\
      \midrule
      \multirow{3}{*}{200}
                                       & FNR                                          & \textbf{8.57}   & 9.60           & 9.68           & 9.31           & 10.00         & 10.86          & 10.54          & \textbf{8.24}                                               & 9.58            \\
                                       & FPR                                          & 3.96            & 1.93           & 2.01           & 2.11           & 1.91          & 1.85           & \textbf{0.20}  & 2.00                                                        & \textbf{0.81}   \\
                                       & F1                                           & 90.88           & 93.00          & 92.80          & 92.74          & 92.91         & 92.75          & 93.07          & \textbf{93.27}                                              & \textbf{95.26}  \\
      \midrule
      \multirow{3}{*}{400}
                                       & FNR                                          & 5.01            & 6.14           & 9.43           & 9.07           & 9.72          & 8.31           & 9.96          & \textbf{4.94}                                               & \textbf{4.08}   \\
                                       & FPR                                          & 3.43            & 1.75           & 2.21           & 2.08           & 1.90          & \textbf{1.69}  & \textbf{0.66}  & 1.95                                                        & 1.88            \\
                                       & F1                                           & 92.25           & 94.63          & 92.64          & 92.88          & 93.03         & 94.10          & \textbf{94.81} & 94.64                                                       & \textbf{95.83}  \\
      \bottomrule
    \end{tabular}%
  }
  \begin{tablenotes}
    \item[] \textsuperscript{†} \model variant using Drebin features (MLP) instead of graph features, with the same adaptation pipeline.
  \end{tablenotes}
\end{table}

\subsection{Robustness to CFG Degradation}
\label{sec:cfg_perturb}

To test whether \model's FCG-level encoder compensates when CFG features degrade, we simulate inference-time obfuscation (e.g., control-flow flattening) by randomly zeroing the features of $p\%$ of CFG nodes. Since zeroing removes a node's signal entirely rather than partially obscuring it, this proxy upper-bounds the impact of any CFG-level obfuscation at the same rate. Pretraining and adaptation are unchanged; we sweep $p$ from $0\%$ to $50\%$ across Task~A's 35 monthly windows.

\begin{table}[t!]
  \centering
  \caption{Robustness of \model to test-time CFG corruption. Mean $\pm$ std over 35 monthly windows with budget $=100$.}
  \label{tab:cfg_perturb}
  \resizebox{\linewidth}{!}{%
    \begin{tabular}{lcccccc}
      \toprule
      $p$ (\%) & 0                  & 10                 & 20                 & 30                 & 40                 & 50                 \\
      \midrule
      F1       & 0.948 $\pm$ 0.016  & 0.944 $\pm$ 0.016  & 0.937 $\pm$ 0.022  & 0.929 $\pm$ 0.026  & 0.920 $\pm$ 0.026  & 0.907 $\pm$ 0.025  \\
      FNR      & 0.395 $\pm$ 0.124  & 0.400 $\pm$ 0.127  & 0.428 $\pm$ 0.132  & 0.475 $\pm$ 0.135  & 0.547 $\pm$ 0.142  & 0.645 $\pm$ 0.136  \\
      FPR      & 0.006 $\pm$ 0.004  & 0.010 $\pm$ 0.011  & 0.014 $\pm$ 0.028  & 0.016 $\pm$ 0.036  & 0.014 $\pm$ 0.031  & 0.010 $\pm$ 0.024  \\
      \bottomrule
    \end{tabular}%
  }
\end{table}

\begin{figure}[t!]
  \centering
  \subfloat[F1 score vs.\ perturbation rate]{%
    \includegraphics[width=0.48\linewidth]{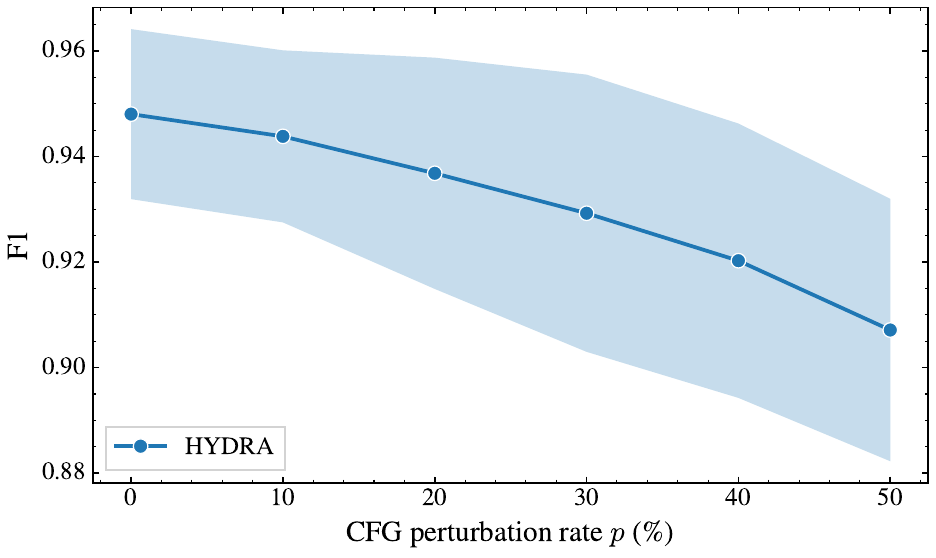}%
    \label{fig:cfg_perturb_f1}}%
  \hfill
  \subfloat[FNR vs.\ perturbation rate]{%
    \includegraphics[width=0.48\linewidth]{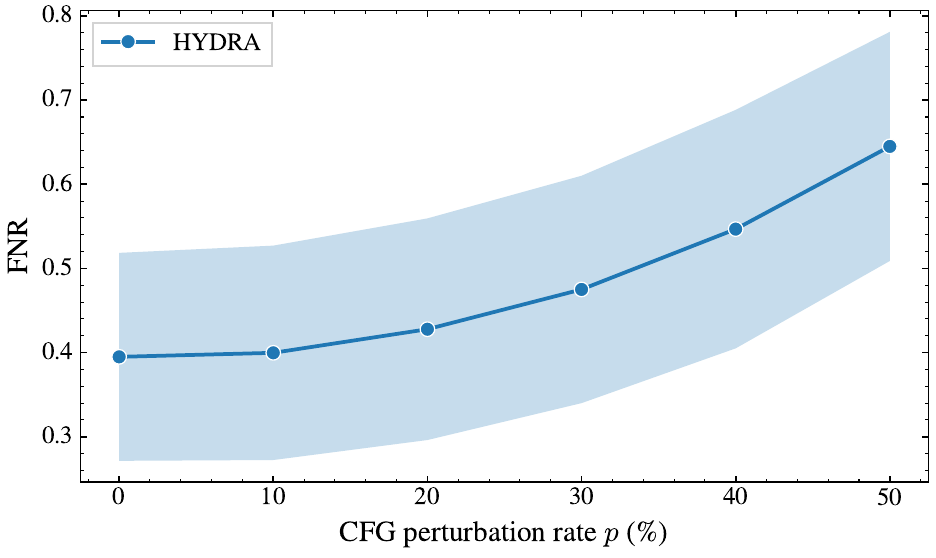}%
    \label{fig:cfg_perturb_fnr}}
  \caption{Effect of test-time CFG corruption on \model. Solid line = mean over 35 monthly windows; shaded band = $\pm 1$ std.}
  \label{fig:cfg_perturb}
\end{figure}

\noindent\textbf{Graceful Degradation under CFG Corruption.}
As presented in Table~\ref{tab:cfg_perturb} and Figure~\ref{fig:cfg_perturb}, \model degrades gracefully rather than collapsing under increasing CFG corruption. Even at $p=50\%$, F1 declines by only 4.1 absolute points (0.948 to 0.907) and FPR stays below 1.6\%, while FNR rises from 0.395 to 0.645 as the model becomes more conservative on the malicious class when intra-procedural signal weakens. Retaining 90.7\% F1 with half the CFG nodes corrupted confirms the hierarchical fall-back: the FCG-level encoder compensates for the lost intra-procedural signal.

\subsection{Inter-Class Drift: Hold-Out Families}
\label{sec:evaluation}

To complement the real-world, time-evolving drift study in Section~\ref{sec:real_world_drift}, we further evaluate \model's resilience to inter-class drift using a hold-out protocol. This setting isolates the effect of previously unseen malware families by holding out one family from training in each fold, allowing it to appear only during adaptation and testing. This leave-one-family-out approach mirrors standard practices in prior work on drift and domain adaptation~\cite{yangCADEDetectingExplaining2021, liRevisitingConceptDrift2025}.

\subsubsection{Experimental Setup}
\label{sec:holdout_setup}
In each evaluation fold, we designate the held-out malware family as the unseen \textit{target domain}, while the remaining families constitute the \textit{source domain}. Benign samples are partitioned across both domains to maintain a realistic class distribution. We repeat this process for each family and report the averaged results.

\textbf{Data Splits and Labeling Budget.}
We construct the source and target domains under a strict class balance constraint: a \textasciitilde1:9 malware-to-benign ratio is maintained in all splits to mitigate spatial bias.

The source domain is split 80/20 into training and testing sets. The target domain is split 50/50 into an adaptation set and a final evaluation set. To simulate realistic analyst constraints, we grant our model access to only a small, budgeted subset of labeled samples from the adaptation set for updates. The remaining adaptation samples are available as unlabeled data. For the \datasetDrebin subset, we test with budgets of 10, 20, 30, 50, and 100 labeled samples; for the \dataset subset, we use 50, 100, 150, 250, and 500. Final performance is measured on the target evaluation set, which remains entirely unseen during training and adaptation.

\subsubsection{Datasets}

\paragraph{\datasetDrebin (Hold-Out Subset).}
We create a subset from \datasetDrebin~\cite{arp2014drebin} by selecting eight malware families with at least 100 samples each, totaling \num{3317} malicious apps. We then add benign applications from AndroZoo~\cite{allix2016androzoo} to construct the dataset. 

\paragraph{\dataset (Hold-Out Subset).}
From the large-scale \dataset corpus (Section~\ref{sec:real_world_drift}), we select the five most frequent malware families. Table~\ref{tab:holdout_families} lists the selected families for both datasets.

\begin{table}[t]
  \centering
  \small
  \caption{Hold-out subsets for leave-one-family-out evaluation. Left: Drebin families ($\geq$ 100 samples). Right: \dataset top-5 families.}
  \begin{tabular}{l r @{\hspace{2em}} l r}
    \toprule
    \multicolumn{2}{c}{\textbf{\datasetDrebin}} & \multicolumn{2}{c}{\textbf{\dataset}}                                          \\
    \cmidrule(lr){1-2} \cmidrule(lr){3-4}
    \textbf{Family}                             & \textbf{\#Samples}                    & \textbf{Family} & \textbf{\#Samples}   \\
    \midrule
    FakeInstaller                               & 925                                   & smsreg          & \num{5020}           \\
    DroidKungFu                                 & 667                                   & dowgin          & \num{3757}           \\
    Plankton                                    & 625                                   & kuguo           & \num{2746}           \\
    GingerMaster                                & 339                                   & ewind           & \num{2398}           \\
    BaseBridge                                  & 330                                   & airpush         & \num{1849}           \\
    Iconosys                                    & 152                                   &                 &                      \\
    Kmin                                        & 147                                   &                 &                      \\
    FakeDoc                                     & 132                                   &                 &                      \\
    \midrule
    \textbf{Total}                              & \textbf{\num{3317}}                   & \textbf{Total}  & \textbf{\num{15770}} \\
    \bottomrule
  \end{tabular}
  \label{tab:holdout_families}
\end{table}

\subsubsection{Implementation Details}
We follow the common graph construction and feature extraction protocol detailed in Section~\ref{sec:real_drift_setup}. For graph-based models, we use the hierarchical graphs (FCGs and CFGs), and for classic models (SVM and MLP), we use the Drebin static feature set~\cite{arp2014drebin}. All reported results are the mean and standard deviation of three independent runs with fixed random seeds to ensure stability.

\subsubsection{Baseline Methods}
We compare \model against representative baselines from three categories using different strategies. To ensure a fair and rigorous comparison, all methods adhere to the identical hold-out protocol, utilizing the same data splits and labeled target budgets. To ensure fair comparison regarding feature representations, classic models (SVM, MLP) use the standard Drebin feature set, while all graph-based models (including ours and ADDA~\cite{liRevisitingConceptDrift2025}) operate on the same hierarchical graph structures.

\noindent\textbf{Cold-Start Learning.} This strategy trains a model from scratch using only the small, budgeted set of labeled target samples, establishing a performance lower bound~\cite{chenContinuousLearningAndroid2023}. We evaluate two such models: a standard Support Vector Machine (\textbf{SVM-Cold}) and a Multi-Layer Perceptron (\textbf{MLP-Cold} ).

\noindent\textbf{Warm-Start Learning.} This approach first trains a model on the source domain and then updates it with the budgeted labeled target samples~\cite{chenContinuousLearningAndroid2023}. We include two variants: \textbf{SVM-Warm}, which trains on the combined source and labeled target data (joint training), and \textbf{MLP-Warm}, where a source-pretrained MLP is fine-tuned on the labeled target samples.

\noindent\textbf{Domain Adaptation.} We also compare against \textbf{ADDA}~\cite{liRevisitingConceptDrift2025}, a state-of-the-art graph-based domain adaptation method previously described in Section~\ref{sec:real_drift_setup}.

\subsubsection{Results}
\label{sec:drift_adaptation}
\begin{table*}[t]
  \centering
  \footnotesize
  \caption{Drift adaptation results on the \dataset and \datasetDrebin datasets. The best and second-best performing baselines for each metric are highlighted in bold. We show the relative improvement of our method, \model.}

  \begin{tabular}{@{}ccc cc cc cc@{}}
    \toprule
    \multirow{2}{*}{\textbf{Dataset}} & \multirow{2}{*}{\textbf{Budget}} & \multirow{2}{*}{\textbf{Metric}} & \multicolumn{2}{c}{\textbf{COLD}} & \multicolumn{2}{c}{\textbf{WARM}} & \multirow{2}{*}{\textbf{\adda}} & \multirow{2}{*}{\textbf{(Ours)~\model}}                                                                                                                    \\
    \cmidrule(lr){4-5} \cmidrule(lr){6-7}
                                      &                                  &                                  & \textbf{MLP}                      & \textbf{SVM}                      & \textbf{MLP}                    & \textbf{SVM}                            &                     &                                                                                            \\
    \midrule
    \multirow{15}{*}{\dataset}        & \multirow{3}{*}{50}              & F1                               & 62.18±2.15                        & 39.66±10.37                       & 89.02±1.68                      & \textbf{89.08±0.80}                     & \textbf{93.05±1.94} & \textbf{93.78±0.40} (\textcolor{blue}{$\uparrow$0.8\%}/\textcolor{blue}{$\uparrow$5.3\%})  \\
                                      &                                  & FPR                              & 36.30±17.18                       & 19.40±27.44                       & 16.70±2.11                      & \textbf{8.44±1.28}                      & \textbf{4.11±2.27}  & \textbf{4.31±1.07} (\textcolor{red}{$\downarrow$4.9\%}/\textcolor{blue}{$\uparrow$48.9\%}) \\
                                      &                                  & FNR                              & 36.55±15.72                       & 76.90±32.67                       & \textbf{5.65±0.98}              & \textbf{7.37±0.39}                      & 9.64±4.86           & \textbf{4.10±0.53} (\textcolor{blue}{$\uparrow$27.4\%}/\textcolor{blue}{$\uparrow$44.4\%}) \\
    \cline{2-9}
                                      & \multirow{3}{*}{100}             & F1                               & 81.29±2.59                        & 48.38±22.69                       & \textbf{91.48±1.25}             & 89.11±0.76                              & \textbf{92.87±1.91} & \textbf{95.38±2.70} (\textcolor{blue}{$\uparrow$2.7\%}/\textcolor{blue}{$\uparrow$4.3\%})  \\
                                      &                                  & FPR                              & 18.53±13.29                       & \textbf{3.83±0.46}                & 4.45±0.88                       & 8.37±1.20                               & \textbf{3.67±1.42}  & \textbf{0.67±0.36} (\textcolor{blue}{$\uparrow$81.7\%}/\textcolor{blue}{$\uparrow$82.5\%}) \\
                                      &                                  & FNR                              & 18.83±11.22                       & 79.25±29.34                       & 12.10±1.55                      & \textbf{7.37±0.39}                      & \textbf{10.40±4.86} & \textbf{8.06±5.31} (\textcolor{blue}{$\uparrow$22.5\%}/\textcolor{red}{$\downarrow$9.4\%}) \\
    \cline{2-9}
                                      & \multirow{3}{*}{150}             & F1                               & 83.99±2.86                        & 44.94±17.83                       & \textbf{91.68±1.12}             & 89.13±0.57                              & \textbf{94.28±1.41} & \textbf{96.58±0.30} (\textcolor{blue}{$\uparrow$2.4\%}/\textcolor{blue}{$\uparrow$5.3\%})  \\
                                      &                                  & FPR                              & 19.05±7.44                        & \textbf{3.32±4.70}                & \textbf{3.11±0.52}              & 8.25±0.89                               & 4.37±1.10           & \textbf{2.67±0.48} (\textcolor{blue}{$\uparrow$14.1\%}/\textcolor{blue}{$\uparrow$19.6\%}) \\
                                      &                                  & FNR                              & 13.05±2.66                        & 82.85±24.25                       & 14.72±1.89                      & \textbf{7.43±0.30}                      & \textbf{7.00±1.88}  & \textbf{4.03±0.82} (\textcolor{blue}{$\uparrow$42.4\%}/\textcolor{blue}{$\uparrow$45.8\%}) \\
    \cline{2-9}
                                      & \multirow{3}{*}{250}             & F1                               & 83.68±1.88                        & 50.05±25.05                       & \textbf{92.18±0.98}             & 89.11±0.68                              & \textbf{93.88±1.07} & \textbf{96.88±0.20} (\textcolor{blue}{$\uparrow$3.2\%}/\textcolor{blue}{$\uparrow$5.1\%})  \\
                                      &                                  & FPR                              & 16.72±7.32                        & 6.97±9.85                         & \textbf{4.11±0.47}              & 8.31±1.13                               & \textbf{5.26±2.10}  & \textbf{2.90±0.83} (\textcolor{blue}{$\uparrow$29.4\%}/\textcolor{blue}{$\uparrow$44.9\%}) \\
                                      &                                  & FNR                              & 15.42±9.87                        & 69.23±43.52                       & 13.91±1.62                      & \textbf{7.43±0.30}                      & \textbf{6.87±2.23}  & \textbf{3.23±0.33} (\textcolor{blue}{$\uparrow$53.0\%}/\textcolor{blue}{$\uparrow$56.5\%}) \\
    \cline{2-9}
                                      & \multirow{3}{*}{500}             & F1                               & 88.41±0.39                        & 38.22±6.28                        & \textbf{95.08±0.58}             & 89.24±0.57                              & \textbf{94.78±0.25} & \textbf{97.08±0.40} (\textcolor{blue}{$\uparrow$2.1\%}/\textcolor{blue}{$\uparrow$2.4\%})  \\
                                      &                                  & FPR                              & 12.96±1.40                        & 33.33±47.14                       & \textbf{4.01±0.73}              & 8.11±0.92                               & \textbf{5.42±2.14}  & \textbf{1.86±0.58} (\textcolor{blue}{$\uparrow$53.6\%}/\textcolor{blue}{$\uparrow$65.7\%}) \\
                                      &                                  & FNR                              & 10.17±0.74                        & 60.71±43.11                       & \textbf{5.65±0.81}              & 7.36±0.31                               & \textbf{4.92±2.44}  & \textbf{3.83±0.86} (\textcolor{blue}{$\uparrow$22.1\%}/\textcolor{blue}{$\uparrow$32.2\%}) \\
    \midrule
    \multirow{15}{*}{\datasetDrebin}  & \multirow{3}{*}{10}              & F1                               & 73.71±1.03                        & 68.10±1.20                        & \textbf{92.88±1.25}             & 91.90±1.22                              & \textbf{91.98±3.40} & \textbf{94.37±0.26} (\textcolor{blue}{$\uparrow$1.6\%}/\textcolor{blue}{$\uparrow$2.6\%})  \\
                                      &                                  & FPR                              & 15.68±1.96                        & 18.10±2.05                        & \textbf{5.15±0.95}              & 6.00±1.12                               & \textbf{5.27±4.62}  & \textbf{5.05±1.73} (\textcolor{blue}{$\uparrow$1.9\%}/\textcolor{blue}{$\uparrow$4.2\%})   \\
                                      &                                  & FNR                              & 25.22±2.76                        & 30.00±3.10                        & \textbf{8.10±1.50}              & 9.12±1.64                               & \textbf{6.12±2.56}  & \textbf{6.12±1.07} (\textcolor{blue}{$\uparrow$0.0\%}/\textcolor{blue}{$\uparrow$24.4\%})  \\
    \cline{2-9}
                                      & \multirow{3}{*}{20}              & F1                               & 76.01±1.50                        & 70.10±1.20                        & \textbf{93.98±1.15}             & 92.90±1.21                              & \textbf{94.28±1.10} & \textbf{94.86±1.10} (\textcolor{blue}{$\uparrow$0.6\%}/\textcolor{blue}{$\uparrow$1.8\%})  \\
                                      &                                  & FPR                              & 12.32±1.20                        & 16.20±1.80                        & \textbf{4.50±0.80}              & 5.50±1.00                               & \textbf{5.27±4.62}  & \textbf{4.30±0.95} (\textcolor{blue}{$\uparrow$4.4\%}/\textcolor{blue}{$\uparrow$18.4\%})  \\
                                      &                                  & FNR                              & 22.24±2.50                        & 28.00±2.90                        & \textbf{7.00±1.40}              & 8.00±1.50                               & \textbf{6.12±2.56}  & \textbf{6.12±2.56} (\textcolor{blue}{$\uparrow$0.0\%}/\textcolor{blue}{$\uparrow$12.6\%})  \\
    \cline{2-9}
                                      & \multirow{3}{*}{30}              & F1                               & 78.28±2.00                        & 72.10±1.20                        & \textbf{94.88±1.05}             & \textbf{93.90±1.22}                     & 93.78±0.50          & \textbf{97.76±0.54} (\textcolor{blue}{$\uparrow$3.0\%}/\textcolor{blue}{$\uparrow$4.1\%})  \\
                                      &                                  & FPR                              & 10.95±1.50                        & 14.15±1.70                        & \textbf{3.80±0.70}              & 4.80±0.90                               & \textbf{4.60±2.13}  & \textbf{3.60±0.80} (\textcolor{blue}{$\uparrow$5.3\%}/\textcolor{blue}{$\uparrow$21.7\%})  \\
                                      &                                  & FNR                              & 19.39±2.40                        & 26.00±2.80                        & \textbf{6.00±1.30}              & \textbf{7.00±1.40}                      & 7.73±2.65           & \textbf{5.73±0.65} (\textcolor{blue}{$\uparrow$4.5\%}/\textcolor{blue}{$\uparrow$18.1\%})  \\
    \cline{2-9}
                                      & \multirow{3}{*}{50}              & F1                               & 80.86±2.30                        & 74.10±1.20                        & \textbf{95.68±0.95}             & \textbf{94.90±1.21}                     & 94.28±0.70          & \textbf{98.22±0.70} (\textcolor{blue}{$\uparrow$2.6\%}/\textcolor{blue}{$\uparrow$3.5\%})  \\
                                      &                                  & FPR                              & 8.22±1.60                         & 12.20±1.60                        & \textbf{3.10±0.60}              & \textbf{4.10±0.80}                      & 5.05±1.87           & \textbf{2.05±0.70} (\textcolor{blue}{$\uparrow$33.9\%}/\textcolor{blue}{$\uparrow$50.0\%}) \\
                                      &                                  & FNR                              & 17.55±2.30                        & 24.00±2.70                        & \textbf{5.00±1.20}              & \textbf{6.00±1.30}                      & 6.38±0.38           & \textbf{4.38±0.38} (\textcolor{blue}{$\uparrow$12.4\%}/\textcolor{blue}{$\uparrow$27.0\%}) \\
    \cline{2-9}
                                      & \multirow{3}{*}{100}             & F1                               & 82.30±0.77                        & 76.10±1.20                        & \textbf{96.28±0.85}             & \textbf{95.60±1.12}                     & 94.38±0.30          & \textbf{98.31±0.36} (\textcolor{blue}{$\uparrow$2.1\%}/\textcolor{blue}{$\uparrow$2.8\%})  \\
                                      &                                  & FPR                              & 6.51±1.11                         & 10.10±1.50                        & \textbf{2.50±0.50}              & \textbf{3.50±0.70}                      & 12.69±8.49          & \textbf{1.69±0.49} (\textcolor{blue}{$\uparrow$32.4\%}/\textcolor{blue}{$\uparrow$51.7\%}) \\
                                      &                                  & FNR                              & 15.18±2.25                        & 22.00±2.60                        & \textbf{4.00±1.10}              & 5.00±1.20                               & \textbf{3.70±1.48}  & \textbf{2.70±1.48} (\textcolor{blue}{$\uparrow$27.0\%}/\textcolor{blue}{$\uparrow$32.5\%}) \\
    \bottomrule
  \end{tabular}
  \label{tab:drift_adaptation}
\end{table*}

\begin{figure*}[t]
  \centering
  \subfloat[F1]{%
    \includegraphics[width=0.25\linewidth]{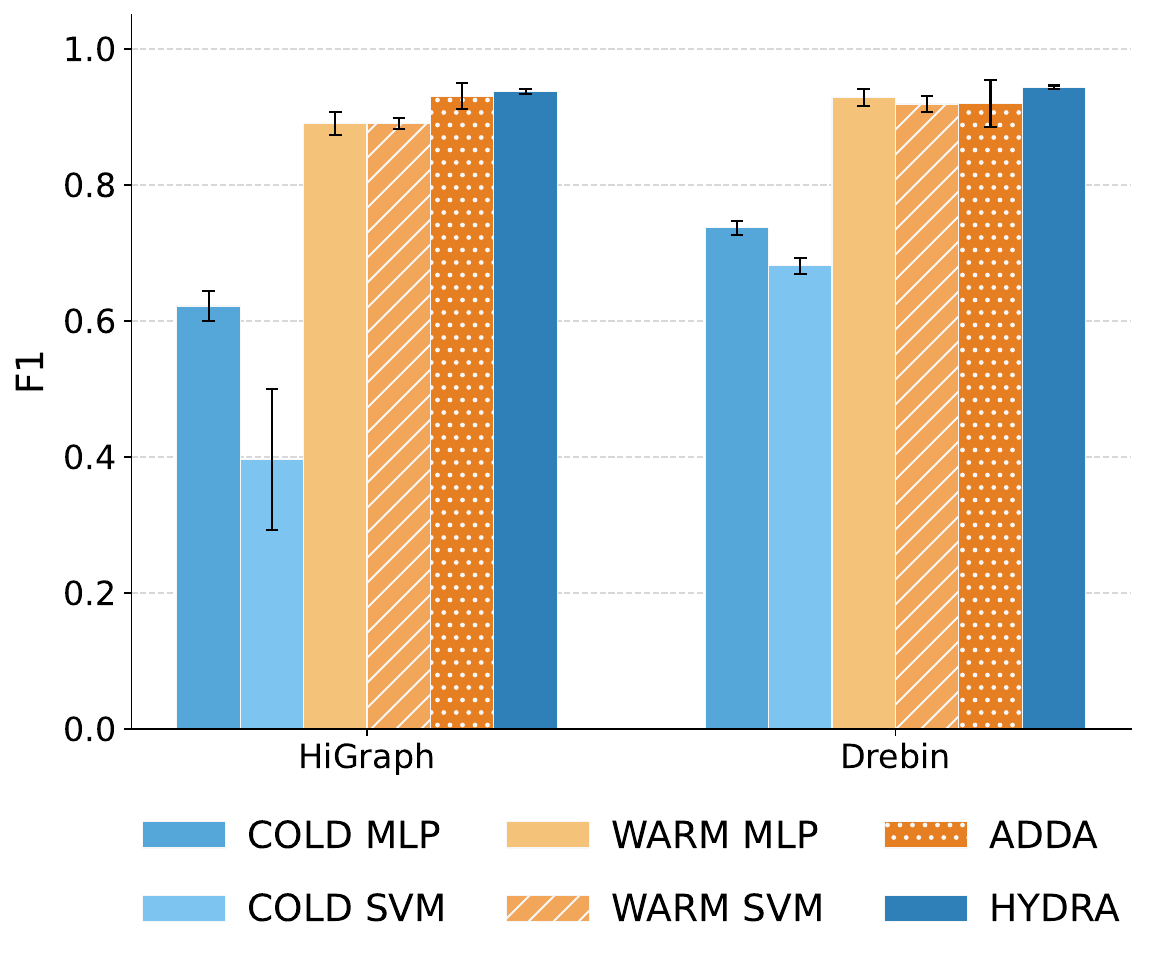}%
    \label{fig:barchart_accuracy}}%
  \hfill
  \subfloat[FNR]{%
    \includegraphics[width=0.25\linewidth]{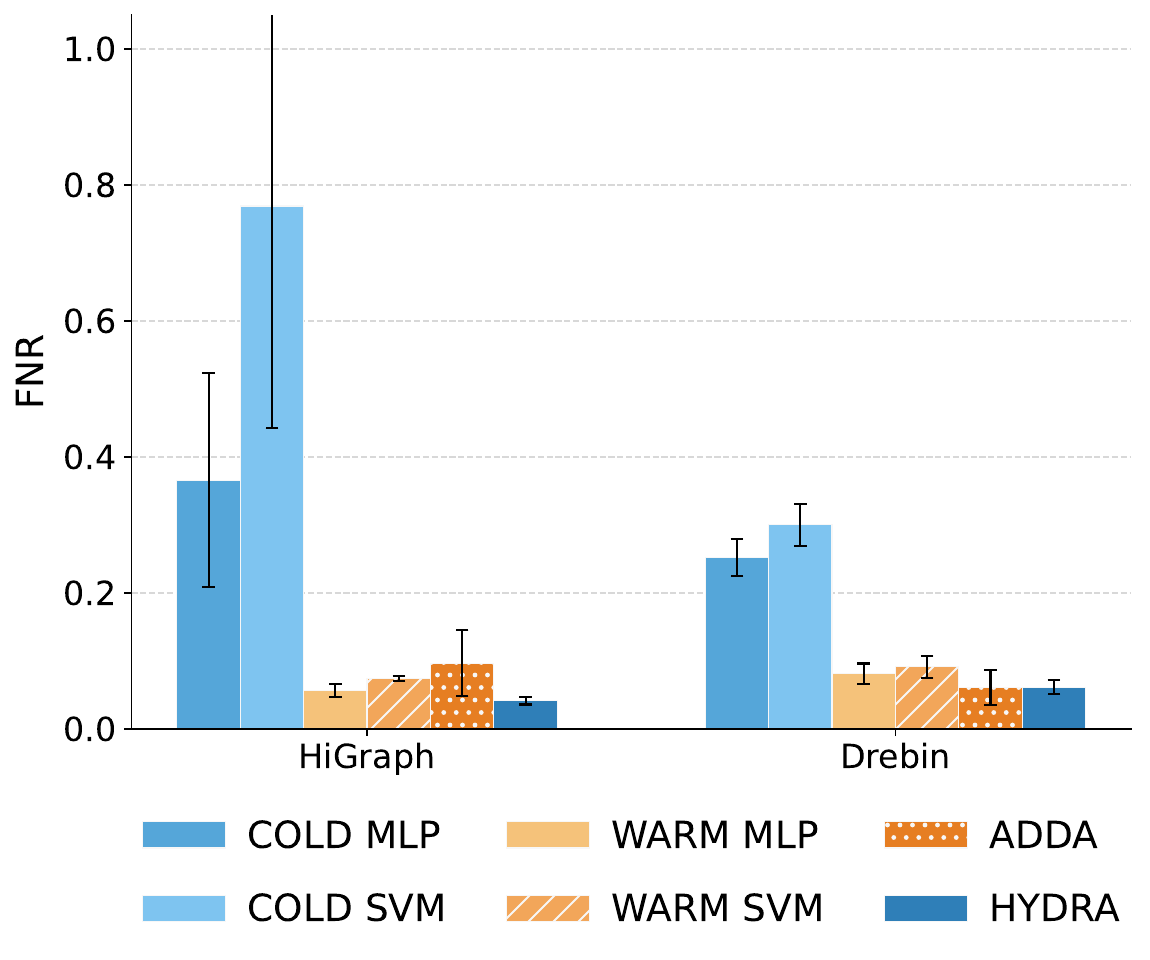}%
    \label{fig:barchart_recall}}%
  \hfill
  \subfloat[F1 vs. Budget]{%
    \includegraphics[width=0.25\linewidth]{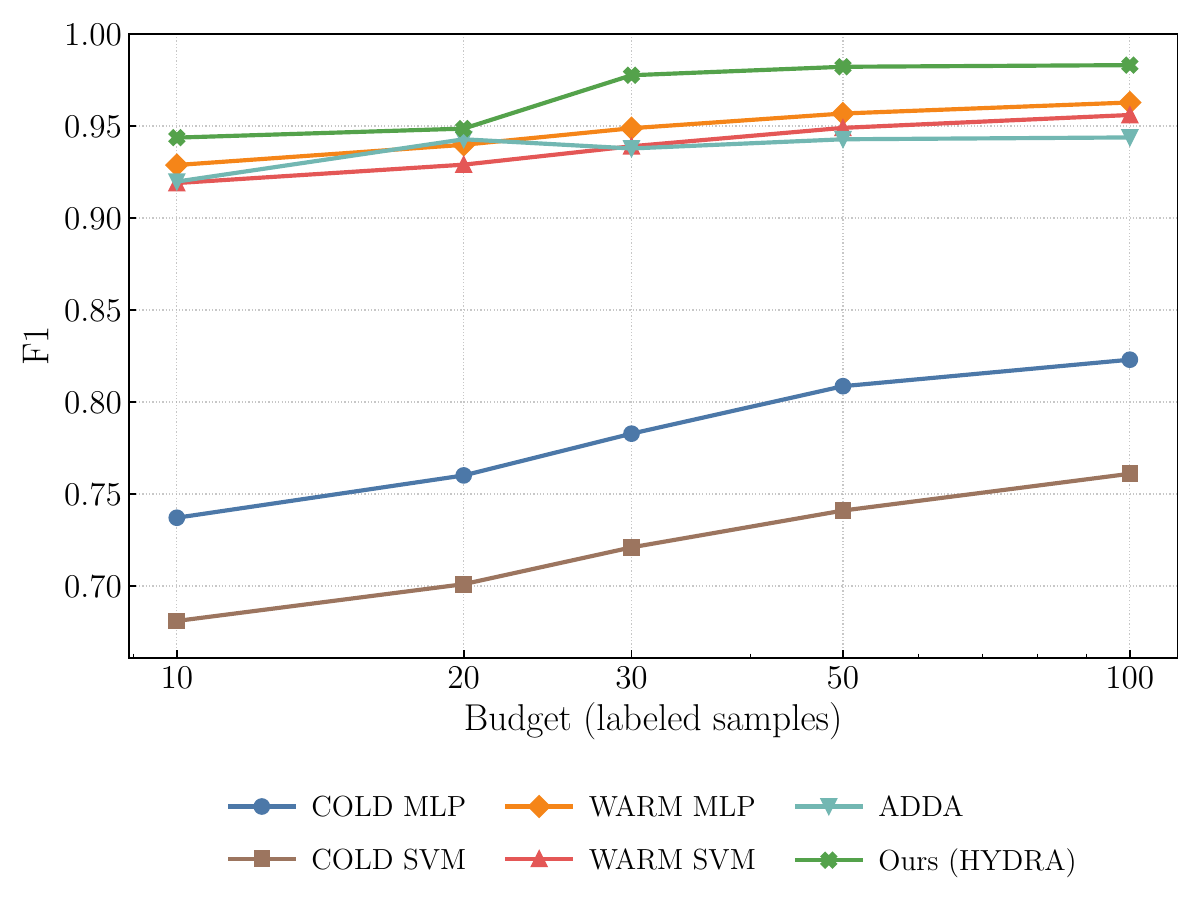}%
    \label{fig:accuracy_trends}}%
  \hfill
  \subfloat[FNR vs. Budget]{%
    \includegraphics[width=0.25\linewidth]{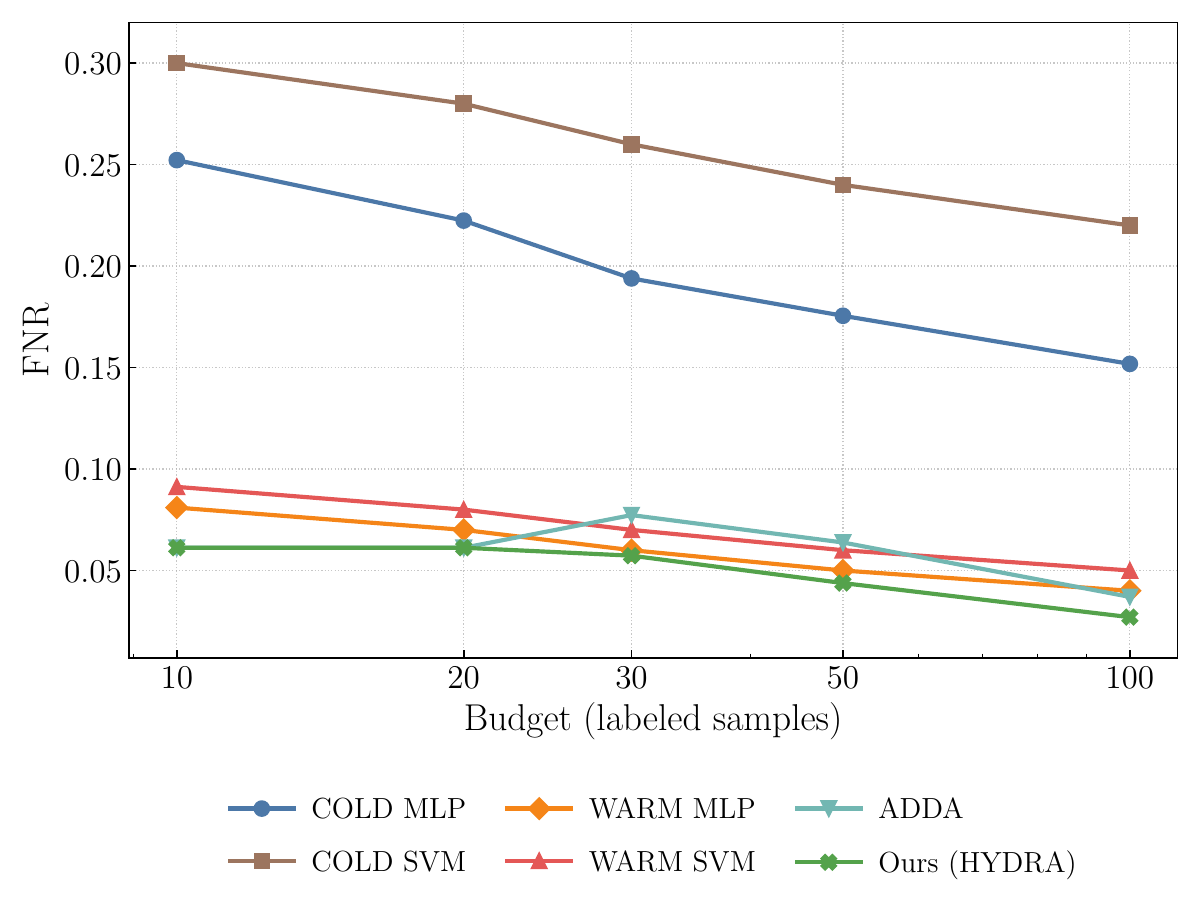}%
    \label{fig:recall_trends}}
  \caption{Performance comparison. (a) and (b) show the adaptation results on the \dataset and \datasetDrebin datasets with a labeling budget of 10 and 50 respectively. (c) and (d) illustrate the performance trends (F1 and FNR) as the budget increases. Our method consistently outperforms baselines, especially in resource-constrained scenarios.}
  \label{fig:performance_trends}
\end{figure*}

Table~\ref{tab:drift_adaptation} and Figure~\ref{fig:performance_trends} summarize our adaptation results. \model consistently and significantly outperforms all cold-start, warm-start, and adversarial domain adaptation (ADDA) baselines on both the \dataset and \datasetDrebin datasets, particularly under stringent labeling budgets.

\noindent \textbf{Performance on Large-Scale Data (\dataset).} On the \dataset dataset, \model demonstrates robust performance and superior adaptation. With a budget of 50 labeled samples, our model achieves an F1-score of 93.78\%, outperforming the strongest warm-start baseline (SVM-Warm, 89.08\%) by +4.70 percentage points and the ADDA baseline (93.05\%) by +0.73 points (Table~\ref{tab:drift_adaptation}). As the budget increases to 500 samples, \model maintains its lead with a 97.08\% F1-score and, critically, reduces the False Positive Rate (FPR) to 1.86\%. This is a 65.7\% relative reduction compared to ADDA's 5.42\% FPR, translating to far fewer false alarms for analysts.

\noindent \textbf{Effectiveness in Low-Resource Scenarios (\datasetDrebin).} The advantages of \model are most pronounced in low-resource settings, a common challenge in security operations. On the \datasetDrebin dataset, with just 10 labeled samples, \model achieves an F1-score of 94.37\%. This is a substantial improvement over cold-start models (which score 73.71\% and 68.10\% F1) and a 1.6\% relative gain over the best warm-start model (92.88\% F1). More critically, our method reduces the False Negative Rate (FNR) to 6.12\%, a 24.4\% relative reduction compared to the best baseline's 8.10\%. This means \model can detect more emerging threats with minimal expert input, a crucial capability for fast-evolving malware landscapes.

Across all scenarios, \model shows superior adaptation performance. Its success stems from the hierarchical graph representation, which captures both high-level and fine-grained behavioral patterns, and the hierarchical contrastive learning objective, which effectively aligns source and target domain distributions. These results confirm that \model provides a more effective and label-efficient solution for adapting malware detectors to concept drift, offering practical benefits by improving detection accuracy, reducing the volume of manual labeling, and lowering false alarm rates.

\subsection{Ablation Study}

We conduct an ablation study to evaluate the contribution of \model's core components: the hierarchical graph structure and the GIN backbone. Table~\ref{tab:ablation_study} presents the results on the \datasetDrebin dataset with a labeling budget of 10 samples.

\noindent \textbf{Hierarchical Structure.} We first evaluate our hierarchical graph representation, which integrates both Function Call Graphs (FCGs) and Control Flow Graphs (CFGs). As shown in Table~\ref{tab:ablation_study}, the full hierarchical model (F1: 94.37\%) outperforms variants using only FCGs (F1: 92.72\%) or only CFGs (F1: 84.08\%). The FCG-only model captures high-level call interactions but misses fine-grained code logic. Conversely, the CFG-only model understands code semantics but lacks a global view of the application, resulting in a high FPR of 14.44\%. Combining both graph levels provides a more comprehensive representation, improving the F1-score by 1.65\% over the FCG-only variant and demonstrating the synergy between global and local structural information.

\noindent \textbf{GNN Backbone.} We also evaluate the impact of the GNN architecture. Table~\ref{tab:ablation_study} shows that GIN (F1: 94.37\%) achieves the best performance. GAT also performs strongly (F1: 93.89\%), leveraging attention to identify important substructures. However, GCN's simpler aggregation scheme is insufficient for modeling complex behaviors, resulting in a lower F1-score of 82.82\%. The superior performance of GIN highlights the importance of its expressive power for distinguishing nuanced graph structures in our contrastive learning setting.

\begin{table}[htbp]
  \centering
  \caption{Ablation study on the \datasetDrebin dataset (Budget=10). We assess variants with different graph structures (FCG-only, CFG-only) and GNN backbones (GCN, GAT).}
  \label{tab:ablation_study}
  \begin{tabular}{@{}llcc@{}}
    \toprule
    \textbf{Ablation}          & \textbf{Method}       & \textbf{F1 Score}   & \textbf{FNR} \\
    \midrule
    \multirow{2}{*}{Structure} & FCG-only              & 92.72±1.15          & 3.24±1.67    \\
                               & CFG-only              & 84.08±2.71          & 16.79±10.31  \\
    \midrule
    \multirow{3}{*}{Backbone}  & \model (GCN)          & 82.82±0.58          & 19.45±7.45   \\
                               & \model (GAT)          & 93.89±0.34          & 6.65±2.27    \\
                               & \textbf{\model (GIN)} & \textbf{94.37±0.26} & 6.12±1.07    \\
    \bottomrule
  \end{tabular}
\end{table}

\subsection{Representation Learning Analysis}

To qualitatively evaluate our method, we visualize the latent representations learned by \model using t-SNE~\cite{van2008visualizing}. We investigate whether our approach produces embeddings that are both \textbf{domain invariant} (bridging the gap between source and target data) and \textbf{class discriminative} (maintaining separation between malware families).

\begin{figure}[t]
  \centering
  \includegraphics[width=\linewidth, trim=0 20 2 0, clip]{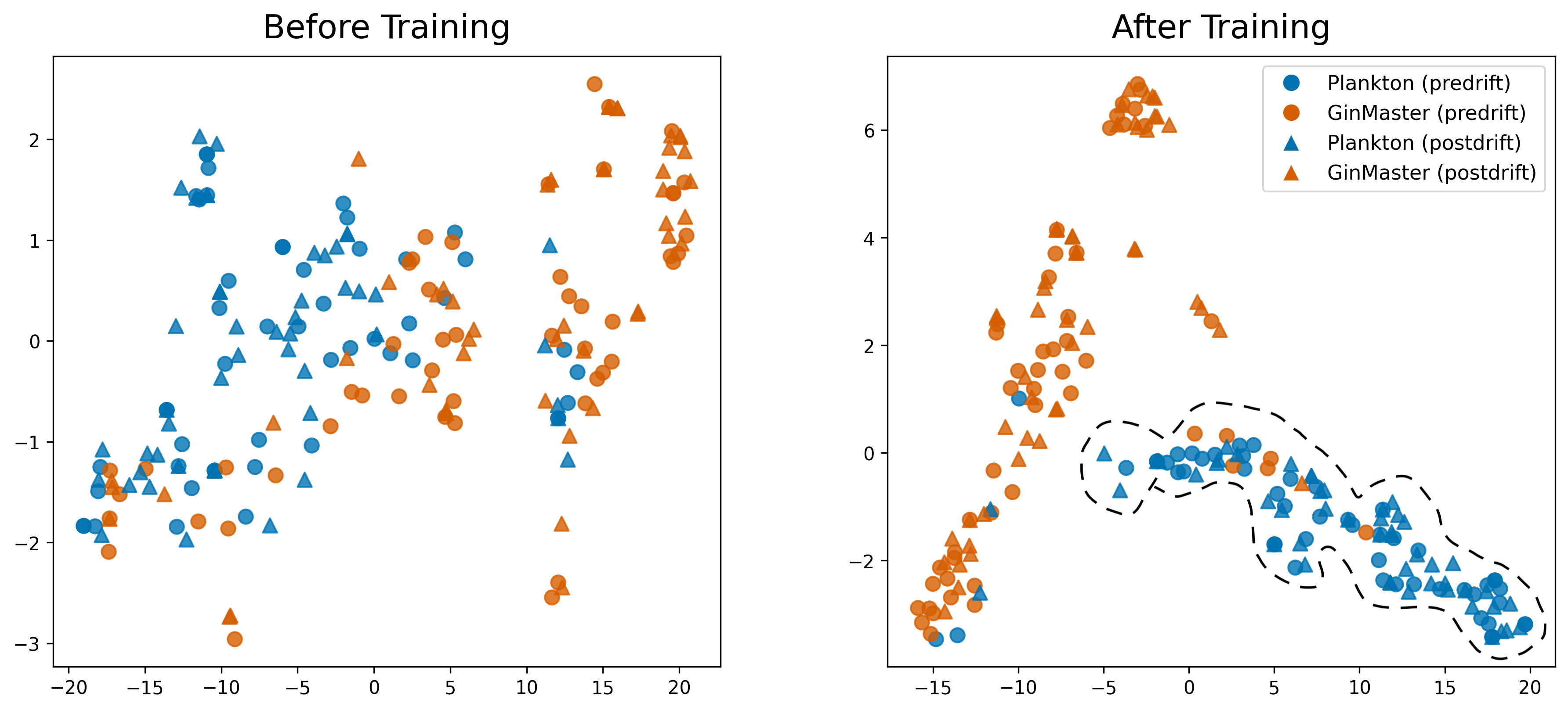}
  \caption{Effectiveness on latent feature distribution. Each point represents a sample. Circles and triangles denote pre-drift and post-drift samples, respectively. (Left) Before adaptation, features learned only from predrift data show a clear distribution shift. (Right) After adaptation with \model, samples from the same malware family form cohesive, domain-invariant clusters, demonstrating effective drift mitigation.}
  \label{fig:embeddings_comparison}
\end{figure}

Figure~\ref{fig:embeddings_comparison} illustrates the adaptation effect on embeddings for two malware families from the \datasetDrebin dataset. Circles denote pre-drift samples and triangles denote post-drift samples. Before adaptation (left panel), a model trained only on pre-drift data produces scattered embeddings for post-drift samples, revealing a clear distribution shift between the domains. This divergence highlights the core challenge of concept drift, where a model fails to generalize to new, evolved samples.

After adaptation with \model (right panel), the distributions are aligned. Samples from the same family, from both pre- and post-drift domains, form tight, coherent clusters. For instance, both pre- and post-drift Plankton samples now form a distinct cluster, separate from GinMaster. This visualization indicates that our model learns drift-invariant and discriminative features, which is key to its superior adaptation performance.

\noindent\textbf{Quantitative Explainability Analysis.} To rigorously verify whether the learned representations are truly drift-invariant and responsible for classification, we complement the t-SNE visualization with a quantitative analysis using \emph{GNNExplainer}~\cite{yingGnnexplainerGeneratingExplanations2019}. We quantify the causal role of learned features using two metrics: \emph{Fidelity+} (Sufficiency), measuring if the explanation subgraph $G_{sub}$ alone is sufficient for prediction, and \emph{Fidelity-} (Necessity), measuring if removing $G_{sub}$ changes the prediction. We compare \model against a \textbf{Source-Only Baseline}, an identical GNN trained exclusively on source data. This comparison isolates the impact of our adaptation strategy on feature attribution.

\begin{table}[t]
  \centering
  \caption{Quantitative explainability analysis. \textbf{Fidelity- (Necessity, higher is better)} measures prediction drop when removing the explanation; \textbf{Fidelity+ (Sufficiency, lower is better)} measures error when retaining only the explanation. The high Fidelity- of \model confirms reliance on drift-invariant features, whereas the Baseline relies on spurious correlations.}
  \label{tab:explanation_fidelity}
  \resizebox{\linewidth}{!}{
    \begin{tabular}{l|c|cc|cc}
      \toprule
      \multirow{2}{*}{\textbf{Family}} & \multirow{2}{*}{\textbf{Samples}} & \multicolumn{2}{c|}{\textbf{Fidelity- ($\uparrow$ Necessity)}} & \multicolumn{2}{c}{\textbf{Fidelity+ ($\downarrow$ Sufficiency)}}                                       \\
                                       &                                   & \textbf{Baseline}                                              & \textbf{\model}                                                   & \textbf{Baseline} & \textbf{\model} \\
      \midrule
      \textit{Aggregated Avg.}         & \textit{314}                      & 0.0639                                                         & \textbf{0.2210}                                                   & \textbf{0.4901}   & 0.5702          \\
      \midrule
      Plankton                         & 50                                & 0.0175                                                         & \textbf{0.1098}                                                   & \textbf{0.4050}   & 0.4501          \\
      FakeInstaller                    & 42                                & 0.2048                                                         & \textbf{0.4422}                                                   & 0.2493            & \textbf{0.1805} \\
      Kmin                             & 25                                & 0.0185                                                         & \textbf{0.2525}                                                   & \textbf{0.7081}   & 0.7357          \\
      DroidKungFu                      & 50                                & 0.0213                                                         & \textbf{0.1344}                                                   & \textbf{0.4408}   & 0.7207          \\
      BaseBridge                       & 50                                & 0.0887                                                         & \textbf{0.2880}                                                   & \textbf{0.5854}   & 0.6755          \\
      GinMaster                        & 48                                & 0.0480                                                         & \textbf{0.1272}                                                   & 0.6640            & \textbf{0.6272} \\
      Iconosys                         & 21                                & 0.0274                                                         & \textbf{0.2820}                                                   & \textbf{0.0562}   & 0.3808          \\
      FakeDoc                          & 23                                & 0.0854                                                         & \textbf{0.1320}                                                   & 0.8228            & \textbf{0.8080} \\
      \bottomrule
    \end{tabular}
  }
\end{table}

Table~\ref{tab:explanation_fidelity} reports results across diverse families. A key finding is the significant disparity in \emph{Fidelity-} (Necessity). The Baseline's low Fid- (avg. 0.0639) suggests reliance on spurious correlations, as removing the identified ``important'' features rarely alters the decision. Conversely, \model achieves a substantially higher Fid- (avg. 0.2210), indicating strong reliance on specific structural patterns. For example, in \emph{FakeInstaller}, \model's Fid- is double that of the Baseline (0.4422 vs. 0.2048). Since these explanations are derived from correctly classified samples in the \emph{target} domain, the high necessity confirms that \model successfully identifies and prioritizes drift-invariant malicious behaviors over domain-specific noise.

\subsection{Parameter Sensitivity Analysis}

\begin{figure}[t]
  \centering
  \includegraphics[width=\linewidth]{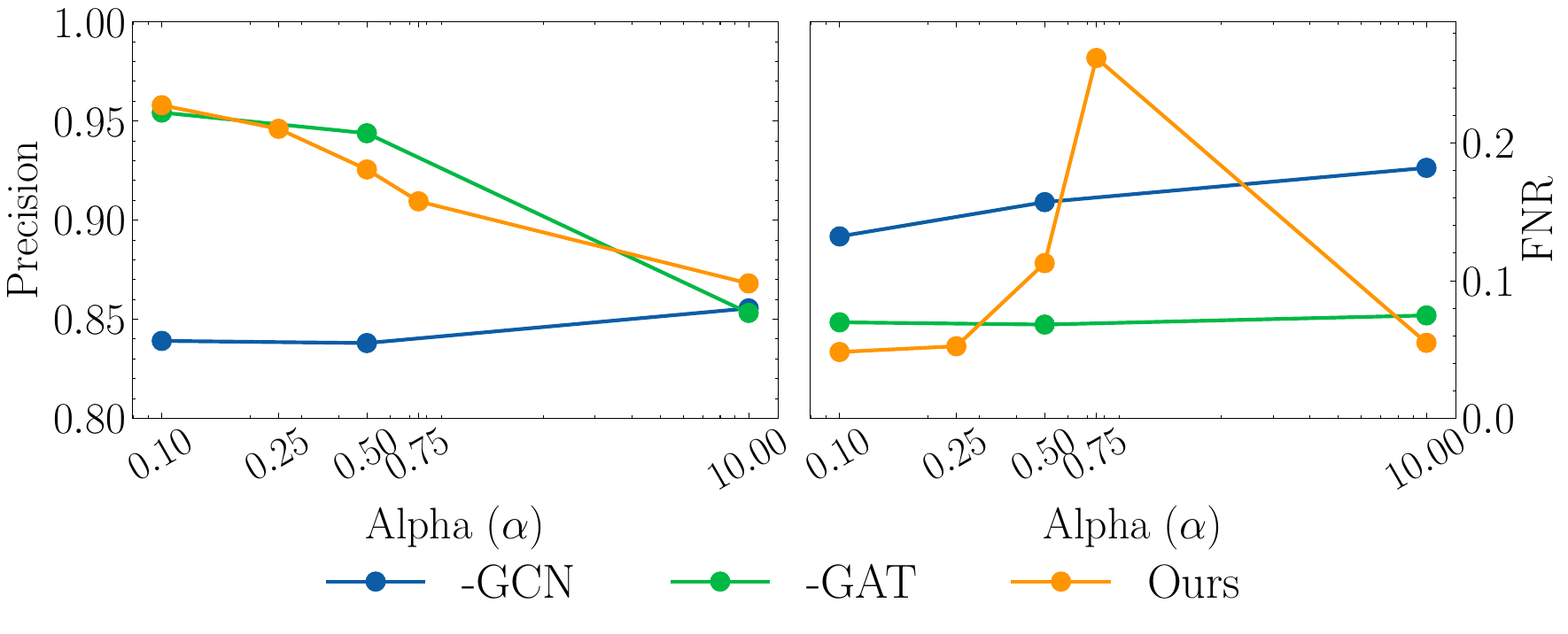}
  \caption{Parameter sensitivity analysis for the loss weight $\alpha$ across different graph neural network backbones. Our \model framework demonstrates strong robustness regardless of the GNN architecture used, though GIN provides the most stable performance.}
  \label{fig:param_sensitivity}
\end{figure}

To evaluate whether the robustness of our framework is independent of a specific GNN architecture, we analyze its sensitivity to the loss weight parameter $\alpha$ using different GNN backbones. We replace the GIN encoder in \model with standard GCN and GAT layers and repeat the experiment. As shown in Figure~\ref{fig:param_sensitivity}, the GIN-based implementation maintains the most stable high precision (>0.90) and low FNR (<0.11) across a wide range of $\alpha$ values. While the GAT and GCN backbones exhibit slightly more performance degradation at extreme $\alpha$ values, they still show considerable robustness. This analysis suggests that the stability of \model is a fundamental property of our joint learning design, rather than an artifact of a single GNN backbone. The optimal balance for our primary model is achieved with $\alpha \in [0.1, 0.75]$.

\section{Discussion}
\label{sec:discussion}

Like all machine learning-based malware detectors, \model is susceptible to evasion attacks. Our static graph construction is vulnerable to obfuscation, and adversaries could manipulate graph structures to poison the contrastive learning process. Defending against such attacks is a significant open challenge. Furthermore, our reliance on pseudo-labeling for continuous adaptation introduces the risk of data poisoning, where an attacker could subtly manipulate the model's decision boundary over time by injecting carefully crafted samples. This risk is inherent to most adaptive security systems and warrants further investigation.

We also note that our proactive domain adaptation framework is orthogonal to sample selection techniques~\cite{liRevisitingConceptDrift2025}. Integrating drift detection to identify minimal divergent subsets as targets for adaptation training could further minimize labeling costs.

Finally, practical deployment would require managing the computational overhead of graph extraction and model updates, likely through techniques such as graph caching, incremental learning, or model distillation.
\section{Conclusion}
\label{sec:conclusion}

In this paper, we introduced \model, a hierarchical graph contrastive learning framework that sets a new direction for handling concept drift in Android malware detection by proactively learning drift-invariant representations. 
Our approach, which jointly models program behavior across hierarchical Control Flow and Function Call Graphs, demonstrates substantial performance gains over existing methods, especially in practical scenarios where labeled data is scarce. The results confirm that combining hierarchical program representations with a tailored cross-domain contrastive learning objective provides a robust defense against evolving threats. This work opens several exciting avenues for future research. Key directions include extending the hierarchical paradigm to new platforms, integrating dynamic analysis to create more comprehensive behavioral models, and advancing the framework towards online continual learning to enable real-time adaptation in production environments. Beyond malware detection, the hierarchical alignment mechanism is essentially task-agnostic and naturally generalises to broader code analysis problems such as software plagiarism detection and cross-architecture vulnerability discovery, where stable structural patterns similarly underlie surface-level code variation. 

\begin{acks}
This work was supported by the National Natural Science Foundation of China
(62266050), the Program for Young and Middle-aged Academic and Technical
Reserve Leaders of Yunnan Province (202205AC160033), the Yunnan Province
Xingdian Talents Support Program, and the Program of Yunnan Key Laboratory of
Intelligent Systems and Computing (202405AV340009). Hanchen Wang is supported
by the Australian Research Council (ARC) under DE250100226. Ying Zhang is
supported by ARC LP210301046. Lu Qin is supported by ARC DP240101322 and
DP260100709.

This paper was edited for grammar and style using Gemini.
\end{acks}

\bibliographystyle{ACM-Reference-Format}
\bibliography{reference}

\section*{Open Science}
Code, configurations, manifests, and preprocessing scripts are available at
\url{https://github.com/hzcheney/HGCL}. Malware samples are withheld because of
licensing restrictions. This version contains the full appendix.

\section*{Ethical Considerations}
We use public malware corpora without human subjects or end-user telemetry and
retain only abstract CFG and FCG structures rather than potentially sensitive
strings. Given the dual-use risk, we release code and configurations but
withhold pretrained weights and functional malware samples.

\appendix
\section{Datasets and Preprocessing}
\label{sec:app-datasets}

We construct time-ordered train/validation/test segments that preserve causal order to reflect real deployment. Labeled target samples are capped by an explicit budget (e.g., 10--500) and selected uniformly at random within each target segment unless otherwise noted; unlabeled data is from the same window. Labels are from prior metadata and not inferred by our model. We decompile each application to extract intra-procedural control-flow graphs (CFGs) and an inter-procedural function-call graph (FCG), removing boilerplate and normalizing identifiers. To evaluate generalization, we hold out specific malware families from training, as detailed in the main text (\autoref{tab:holdout_families}). For \datasetDrebin, we follow established preprocessing, while for \dataset, we strictly adhere to timestamp ordering. All splits are reproducible via manifests in our artifact. A summary of dataset statistics is presented in \autoref{tab:dataset_stats_appendix}.

\begin{table}[ht!]
  \centering
  \caption{Summary statistics for the datasets used in our evaluation.}
  \label{tab:dataset_stats_appendix}
  \begin{tabular}{@{}lr@{\hspace{1.4em}}lr@{}}
  \toprule
  \multicolumn{2}{c}{\textbf{\datasetDrebin}} &
  \multicolumn{2}{c}{\textbf{\dataset}} \\
  \cmidrule(r){1-2}\cmidrule(l){3-4}
  \textbf{Family} & \textbf{\# Samples} & \textbf{Family} & \textbf{\# Samples} \\
  \midrule
  FakeInstaller & 925 & smsreg   & \num{5020} \\
  DroidKungFu   & 667 & dowgin   & \num{3757} \\
  Plankton      & 625 & kuguo    & \num{2746} \\
  GingerMaster  & 339 & ewind    & \num{2398} \\
  BaseBridge    & 330 & airpush  & \num{1849} \\
  Iconosys      & 152 & gappusin & \num{1758} \\
  Kmin          & 147 & adwo     & \num{1081} \\
  FakeDoc       & 132 & youmi    & \num{952}  \\
                &     & hiddad   & \num{885}  \\
                &     & wapron   & \num{860}  \\
  \bottomrule
\end{tabular}

\end{table}

\section{Implementation Details}
\label{sec:app-implementation}

We detail our implementation to facilitate reproduction. Our model encodes per-function CFGs and the application-level FCG using a message-passing GNN (3--5 layers, 128--256 hidden size, ReLU, dropout, and layer normalization). For external API calls lacking CFGs, we embed their names using a frozen pre-trained text model (Qwen3-Embedding-0.6B)~\cite{qwen3embedding}, which are L2-normalized before GNN input. Baselines using FCGs receive identical API embeddings for fair comparison. Per-dataset hyperparameters not specified here are in our supplementary artifact.

We optimize a combined contrastive and classification loss \(\mathcal{L} = \mathcal{L}_{contrast} + \alpha\, \mathcal{L}_{cls}\) using Adam with cosine decay and warmup. The contrastive objective is InfoNCE with temperature \(\tau\), applied to L2-normalized embeddings, with sampling balanced between source and target domains. We use pseudo-labels with a confidence threshold for unlabeled target data, refreshing them periodically to stabilize training. We report F1, FPR, and FNR averaged over at test time, with the operating threshold selected on the validation split. All experimental variables are recorded in configuration files for reproducibility.

\paragraph{Computational Environment} All experiments were conducted on a high-performance computing cluster. Each node is equipped with two Intel Xeon Gold 6346 16-core CPUs, 256GB of DDR4 RAM, and two NVIDIA A40 GPUs, each with 48GB of memory. The storage configuration per node includes a RAID 1 array for the operating system and a high-speed RAID 10 array for data. Our full dataset of APKs requires approximately 18TB of storage space.
\begin{figure*}[t!]
  \centering
  \includegraphics[width=0.9\textwidth]{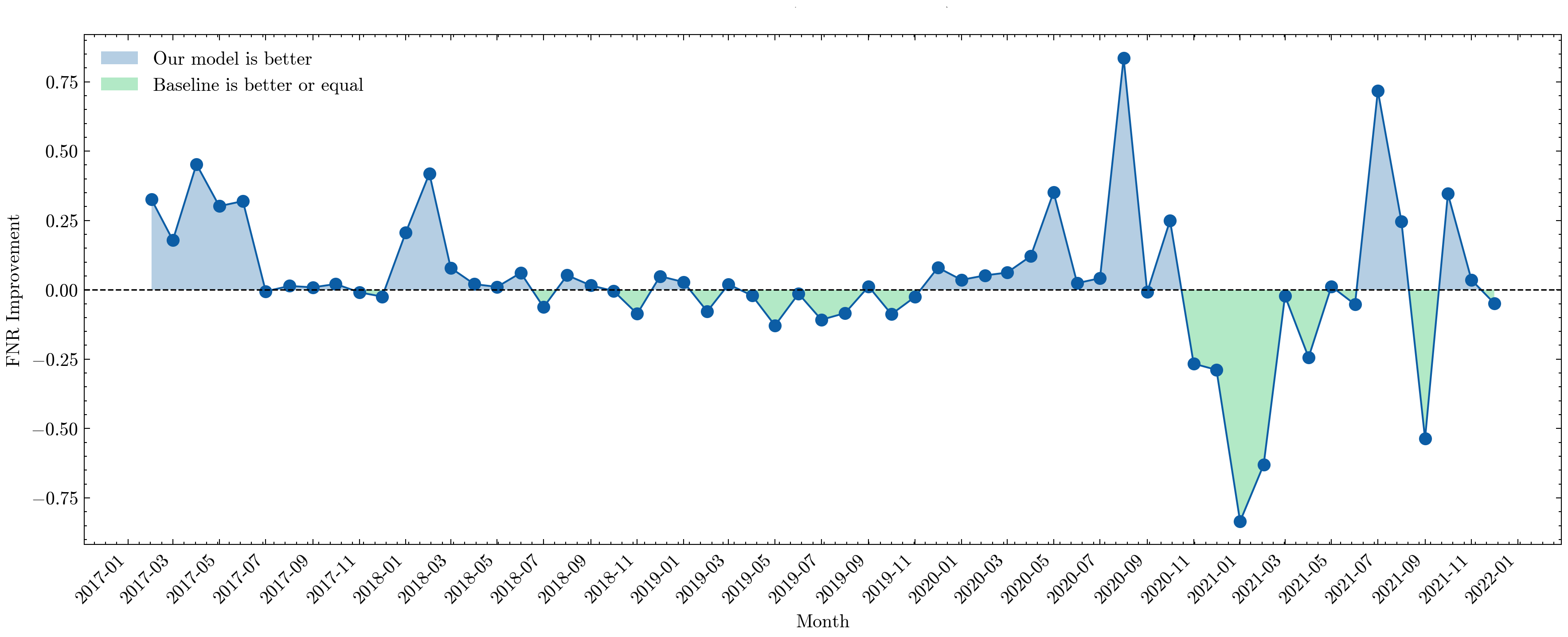}
  \caption{Monthly FNR difference between \model~(Incremental) and HCC at a fixed budget of 200 labels for Task B (2017--2022). Negative values indicate a lower FNR for our method, demonstrating its sustained advantage, particularly during periods of significant concept drift.}
  \label{fig:perf_delta_over_time}
\end{figure*}

\begin{figure*}[t!]
  \centering
  \subfloat[F1 - Adapt vs HCC]
  {\includegraphics[width=0.48\textwidth]{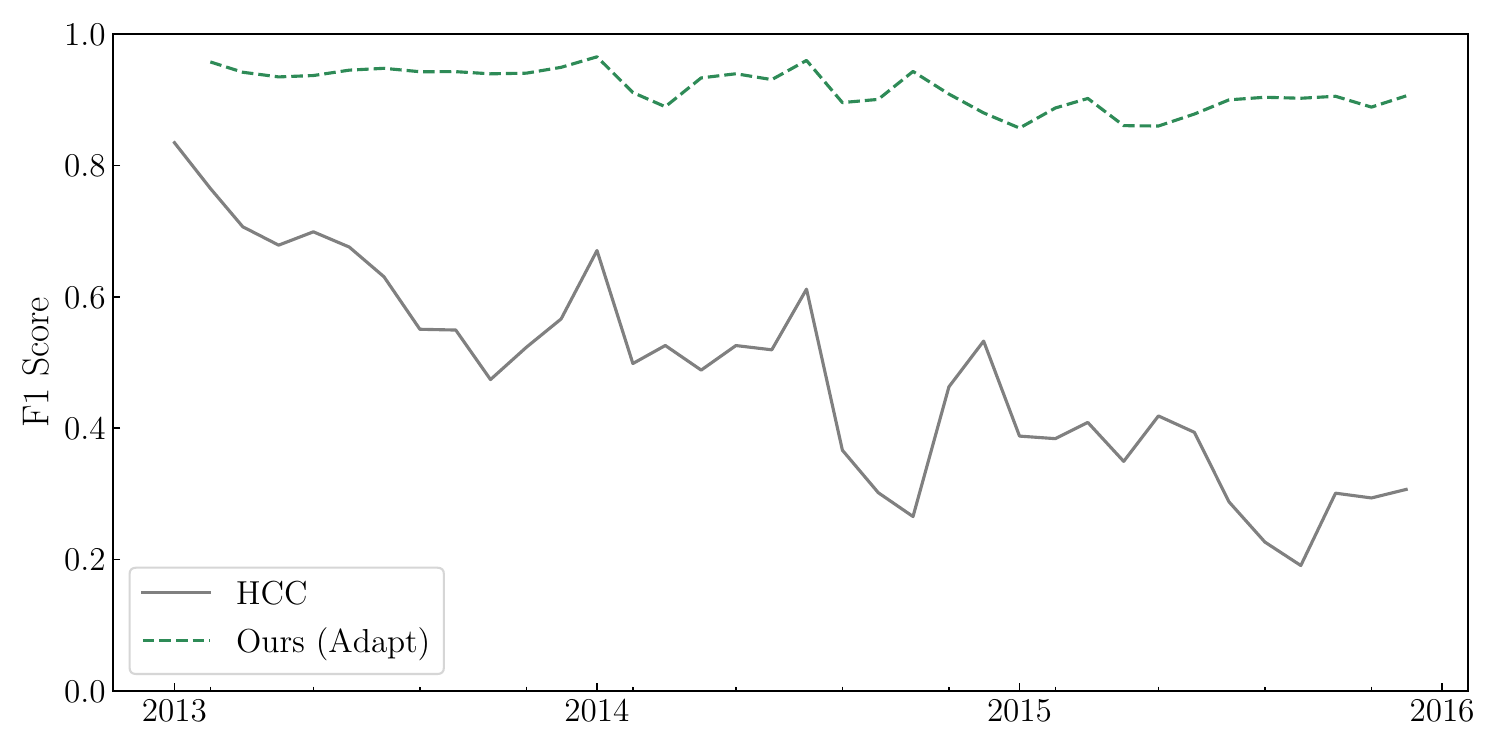}
    \label{fig:f1_adapt_hcc}}
  ~
  \subfloat[FNR - Adapt vs HCC]
  {\includegraphics[width=0.48\textwidth]{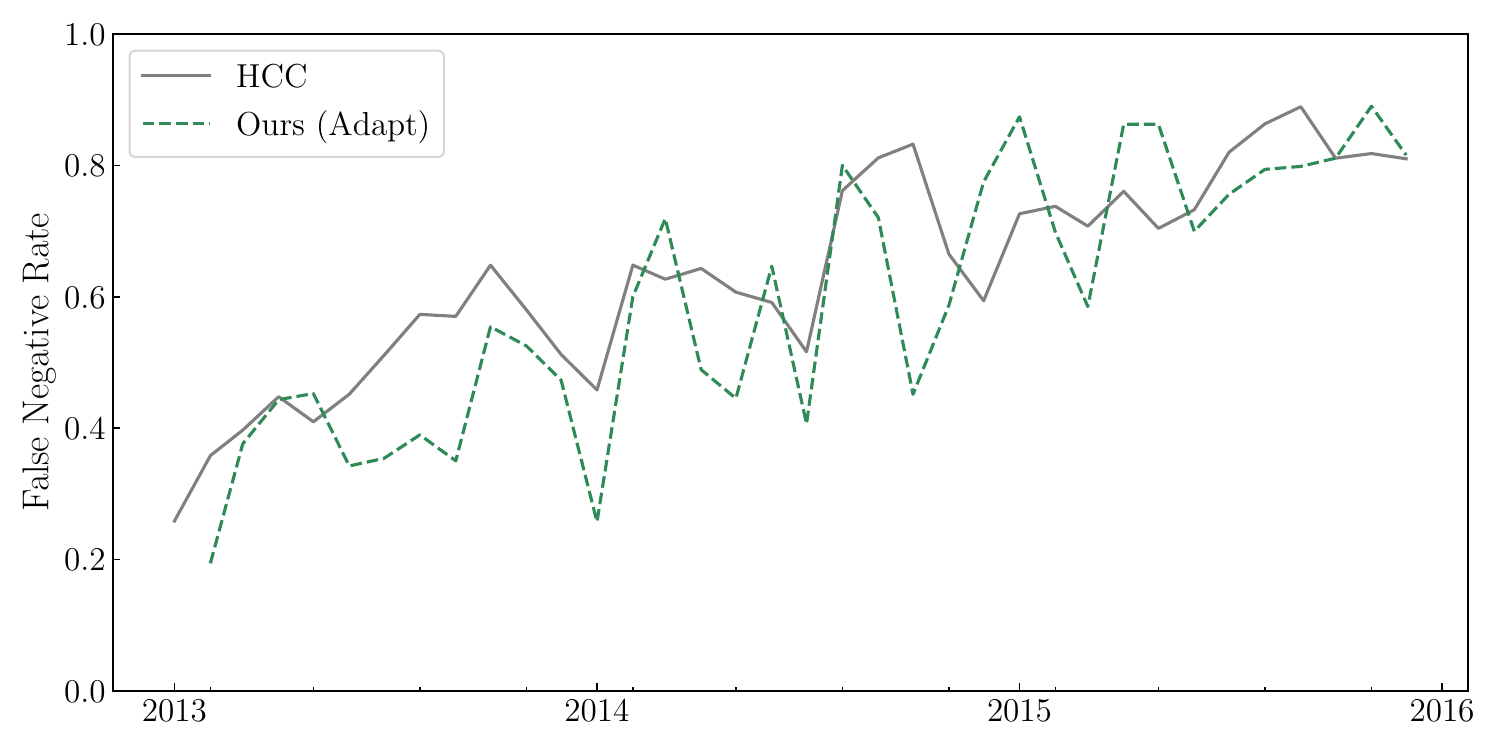}
    \label{fig:fnr_adapt_hcc}}
  \caption{Temporal performance vs. HCC for the Adapt variant. F1 and FNR scores over time.}
  \label{fig:real_drift_hcc_adapt}
\end{figure*}

\paragraph{Hyperparameter Settings} We performed a grid search over key hyperparameters on the validation set for each temporal window. The final configuration used for our model is detailed in \autoref{tab:hyperparams}. For baseline models, we followed standard practices. For SVM, we used an RBF kernel with the regularization parameter \(C\) tuned from \{0.1, 1, 10, 100\}. For MLP-based baselines, we used a 3-layer network with 256 hidden units, ReLU activations, and dropout (0.5). For ADDA, the discriminator was a 3-layer MLP trained with a separate Adam optimizer, and its alignment schedule followed the original paper. For HCC, CADE, and Trans, we utilized their publicly available implementations and followed the authors' recommendations for hyperparameter tuning to ensure a fair comparison.

\begin{table}[ht!]
  \centering
  \caption{Key hyperparameters for our \model~model.}
  \label{tab:hyperparams}
  \begin{tabular}{lc}
    \toprule
    \textbf{Hyperparameter}            & \textbf{Value}         \\
    \midrule
    GNN Backbone                       & GIN                    \\
    GNN Layers                         & (3, 5)                 \\
    Hidden Dimension                   & (128, 256)             \\
    Optimizer                          & Adam                   \\
    Learning Rate                      & (1e-4, 1e-5)           \\
    Batch Size                         & (16, 32, 64, 128)      \\
    Training Epochs                    & (20, 50, 100)          \\
    Early Stopping                     & Patience 10            \\
    Contrastive Temperature (\(\tau\)) & (0.05, 0.1, 0.2)       \\
    Loss Weight (\(\alpha\))           & (0.1, 0.25, 0.5, 0.75) \\
    Pseudo-label Threshold             & 0.9                    \\
    Dropout Rate                       & 0.5                    \\
    \bottomrule
  \end{tabular}
\end{table}

\section{Additional Results}
\label{sec:app-results}

To further illustrate our model's sustained performance advantage under concept drift, we present additional results from our temporal evaluation. \autoref{fig:perf_delta_over_time} plots the monthly difference in False Negative Rate (FNR) between our incremental adaptation approach (\model) and the Hierarchical Co-clustering (HCC) baseline on Task B, using a fixed budget of 200 labels. The negative values consistently show that our model achieves a lower FNR. The performance gap is particularly pronounced during periods where significant concept drift is known to occur (e.g., late 2018 and mid-2020), highlighting our method's ability to effectively adapt to evolving threats over time.

\autoref{fig:real_drift_hcc_adapt} reports the \textit{Adapt} variant against
HCC. \textit{Adapt} retains F1 near 0.9 over most monthly windows, whereas HCC
declines substantially as drift accumulates. Its FNR is more variable and the
curves cross in several months, but performance remains broadly competitive
even though \textit{Adapt} reloads the source-pretrained model rather than
carrying weights forward. These results complement the \textit{Incremental}
evaluation in \autoref{fig:real_drift_hcc_comparison}.

\section{Reproducibility and Artifact}
\label{sec:app-reproducibility}

To ensure full reproducibility, we provide a supplementary artifact containing all source code, configuration files, dataset hashes, and experiment scripts. We also provide a new notebook in the repository root directory for simplified reproduction. To use it, one needs to first download the pre-processed feature data, unzip it, and then run the notebook directly.

Our implementation is built on Python 3.8+ and TensorFlow; we use \texttt{uv} for streamlined dependency management. The artifact is organized with dedicated directories for source code (\texttt{src/}), configurations (\texttt{configs/}), data manifests (\texttt{data/}), and experiment outputs (\texttt{experiments/}).

The end-to-end experimental pipeline consists of four main stages, automated via scripts:
\begin{enumerate}
  \item \textbf{Data Acquisition:} APK samples are downloaded from AndroZoo using their SHA256 hashes via our \path{monthly_malware_downloader.py} script.
  \item \textbf{Graph Extraction:} Downloaded APKs are converted into hierarchical graphs using \path{apk_to_higraph.py}. This script performs static analysis to extract both inter-procedural Function Call Graphs (FCGs) and intra-procedural Control Flow Graphs (CFGs).
  \item \textbf{Graph Preprocessing:} We use \path{higraph-preprocess-graph.py} to process raw graphs into a model-compatible format, generating node feature matrices and sparse adjacency matrices for each sample.
  \item \textbf{Model Training and Evaluation:} The main training script, \texttt{train.py}, orchestrates all experiments. It supports multiple evaluation modes, including static pre-training, monthly adaptation with a fixed budget, and continuous incremental learning. Key parameters such as the model architecture (e.g., GCD-GIN), time windows for training/testing, and labeling budgets are configurable via command-line arguments, allowing for precise replication of our results.
\end{enumerate}

\end{document}